\documentclass[12pt,twoside,fleqn]{article}
\usepackage[eqsecnum]{cmp2e}
\usepackage{amsbsy,amssymb,amstext,epsf}

\newcommand{\be}{\begin{equation}}
\newcommand{\ee}{\end{equation}}
\newcommand{\ba}{\begin{array}}
\newcommand{\ea}{\end{array}}
\newcommand{\bea}{\begin{eqnarray}}
\newcommand{\eea}{\end{eqnarray}}
\newcommand{\bean}{\begin{eqnarray*}}
\newcommand{\eean}{\end{eqnarray*}}

\newcommand{\lp}{\left(}
\newcommand{\rp}{\right)}
\newcommand{\ls}{\left[}
\newcommand{\rs}{\right]}
\newcommand{\lc}{\left\{}
\newcommand{\rc}{\right\}}

\newcommand{\la}{\langle}
\newcommand{\La}{\left\la}
\newcommand{\ra}{\rangle}
\newcommand{\Ra}{\right\ra}

\newcommand{\ket}{\rangle}

\renewcommand{\d}{\mbox{\rm d}}

\newcommand{\gb}%
	{\text{\bf\slshape g%
	}}

\newcommand{\intl}{\int\limits}
\newcommand{\suml}{\sum\limits}
\newcommand{\Imm}{\,\Im{\mathfrak{m}}\,}
\newcommand{\Ree}{\,\Re{\mathfrak{e}}\,}
\newcommand{\im}{{\rm i}}

\newcommand{\veps}{\varepsilon}
\newcommand{\vphi}{\varphi}
\newcommand{\vrho}{\varrho}

\newcommand{\mPhi}{{\mit\Phi}}
\newcommand{\mPsi}{{\mit\Psi}}

\newcommand{\Chi}{\mbox{\parbox[c]{0.8em}{\large$\chi$}}}

\newcommand{\SP}{{\scr P}}

\newcommand{\Sp}{\mathop{\rm Sp}\nolimits}

\newcommand{\const}{\mathop{{\rm const}}\nolimits}

\newcommand{\ddt}{\frac{\partial}{\partial t}}
\newcommand{\dd}[1]{\frac{\partial}{\partial #1}}
\newcommand{\refp}[1]{(\ref{#1})}

\newcommand{\ds}{\displaystyle}

\newcommand{\e}{{\rm e}}
\renewcommand{\SP}{{\wp}}
\newcommand{\vt}{\vartriangle\!\!}
\title[Thermo field dynamics of quantum nuclear systems]
{Thermo field hydrodynamic and kinetic equations of dense quantum nuclear 
systems}

\author[M.V.Tokarchuk, T.Arimitsu, A.E.Kobryn]
{M.V.Tokarchuk\refaddr{a1}, T.Arimitsu\refaddr{a2}, A.E.Kobryn\refaddr{a1}}

\addresses{
\addr{a1}
Institute for Condensed Matter Physics 
of the Ukrainian National Academy of Sciences,
1 Svientsitskii St., UA--290011 Lviv--11, Ukraine
\addr{a2}
Institute of Physics, University of Tsukuba, Ibaraki 305, Japan
}

\date{\today}

\begin{document}
\maketitle

\begin{abstract}
Basic equations of nonequilibrium thermo field dynamics of dense quantum 
systems are presented. A formulation of nonequilibrium thermo field 
dynamics has been performed using the nonequilibrium statistical operator 
method by D.N.Zubarev. Hydrodynamic equations have been obtained in thermo 
field representation. Two levels of the description of kinetics and 
hydrodynamics of a dense nuclear matter are considered. The first one 
is a quantum system with strongly coupled states, the second one is a 
quark-gluon plasma. Generalized transfer equations of a consistent 
description of kinetics and hydrodynamics have been obtained, as well as 
limiting cases are considered.
\pacs 12.38.Mh, 24.85.+p, 52.25.Dg, 52.25.Fi, 82.20.Mj
\end{abstract}


\newcommand{\ph}{\phantom{\int_0^0\bigg[}}
\newcommand{\bmv}[1]{\boldsymbol #1}
\newcommand{\av}{a^{+}}
\newcommand{\ai}{a^{\phantom{+}}}
\newcommand{\hav}{\hat{a}^+}
\newcommand{\tav}{\tilde{a}^+}
\newcommand{\hai}{\hat{a}^{\phantom{+}}}
\newcommand{\tai}{\tilde{a}^{\phantom{+}}}
\newcommand{\hgv}{\hat{\gamma}^+}
\newcommand{\tgv}{\tilde{\gamma}^+}
\newcommand{\hgi}{\hat{\gamma}^{\phantom{+}}}
\newcommand{\tgi}{\tilde{\gamma}^{\phantom{+}}}
\newcommand{\php}{{\phantom B}}
\newcommand{\vrhoq}{\vrho_{\rm q}}
\newcommand{\hvrhoq}{\hat{\vrho}_{\rm q}}

\section{Introduction}


The development of methods for the construction of kinetic and hydrodynamic
equations in the theory of nonequilibrium processes for temperature quantum
field systems is, in particular, important for the 
investigation of nonequilibrium properties of a quark-gluon plasma 
\cite{c1,c2,c3,c4,c5} -- one of the nuclear matter states which can be 
created at ultrarelativistic collisions of heavy nuclei \cite{c6,c7,c8,c9}. 
The description of strongly nonequilibrium processes of a nuclear matter 
appears also in the laser thermonuclear synthesis in systems D-D, D-T, 
D-$^3$He, p-B \cite{c10,c11,c12,c13}. In theoretical investigations of the 
laser thermonuclear synthesis there is a problem of description of a laser 
beam propagation and its absorption by the target (in particular, by the D-T 
mixture), electron and photon energy transport in an ionized magnetized 
plasma with the creation of a corona -- an electron liquid 
($n_e\sim10^{21}\div10^{22}$ cm$^{-3}$), and the creation of a core -- a 
super dense ion liquid ($n\sim10^{24}\div10^{26}$ cm$^{-3}$). In target 
ablation and implosion processes it is important to note transport 
coefficients, such as thermal conductivity, electron's conductivity, as well 
as dielectric properties, generation and evolution of spontaneous magnetic 
fields (with the inductivity of thousands of Tesla), nuclear synthesis 
mechanisms and energy transport by $\alpha$-particles and neutrons which 
are products of synthesis reactions. In this connection, there is a series 
of important problems, when kinetic and hydrodynamic processes in a 
magnetized super dense electron-ion plasma should be described consistently. 
Nonlinear hydrodynamic fluctuations of a magnetized degenerate electron 
liquid in a corona should be considered by taking into account both the 
kinetics of super heat electrons and nonlinear hydrodynamics of a super 
dense magnetized ion liquid, where nuclear synthesis reactions take place 
with the creation of high energy neutrons (with concentration $\sim10^{24}$ 
cm$^{-3}$) and $\alpha$-particles. The kinetics of these particles affects 
the synthesis processes and energy transport in the whole system. From the 
point of view of theoretical methods, in order to describe such strongly 
correlated nonequilibrium processes, one needs to construct the kinetics and 
hydrodynamics of a super dense high temperature electron-ion plasma and a 
nuclear matter at ultrarelativistic collisions of heavy nuclei. A nuclear 
matter -- a stage of a quark-gluon plasma -- appears after the compression 
of the target core of D-D, D-T plasma. As this takes place, its density 
increases by a factor of ten in the fourth degree and the distance between 
nuclons in the centre reaches $\sim 10^{-13}$ cm. Such systems are examples 
of strong both long-range and short-range (nuclear) interactions. There is 
no small parameter for these systems (density, for example). Nonequilibrium 
processes have a strongly correlated collective nature. That is why methods 
which are based on a one-particle description, in particular, on the basis 
of Boltzmann-like kinetic equations, cannot be used. In addition to high 
temperature dense quantum systems, there are Bose and Fermi systems at low 
temperatures with decisive many-particle dissipative correlations. Neither 
the linear response theory nor Boltzmann-like kinetic equations are 
sufficient for their description.

Analysis of the problem of a description of kinetic processes in highly
nonequilibrium and strongly coupled quantum systems on the basis of the
nonequilibrium real-time Green functions technique \cite{c14,c15,c16} and
the theory in terms of non-Markovian kinetic equations describing memory
effects \cite{c17,c18,c19} was made in recent paper \cite{c20} and then in 
monograph \cite{c21}. It is important to note that in \cite{c20} the quantum 
kinetic equation for a dense and strongly coupled nonequilibrium system was 
obtained when the parameters of a shortened description included a 
one-particle Wigner distribution function and an average energy density. On 
the basis of this approach the quantum Enskog kinetic equation was obtained 
in \cite{c21}. This equation is the quantum analogue of the classical one 
within the revised Enskog theory \cite{c22,c23}. Problems of the 
construction of kinetic and hydrodynamic equations for highly nonequilibrium 
and strongly coupled quantum systems were considered based on the 
nonequilibrium thermo field dynamics in \cite{c24,c25,c26,c27,c28,c29}. In 
particular, a generalized kinetic equation for the average value of the 
Klimontovich operator was obtained in \cite{c25} with the help of the 
Kawasaki-Gunton projection operator method. The formalism of the 
nonequilibrium thermo field dynamics was applied to the description of a 
hydrodynamic state of quantum field systems in paper \cite{c26}. Generalized 
transport equations for nonequilibrium quantum systems, specifically for 
kinetic and hydrodynamic stages, were obtained in \cite{c27} on the basis of 
the thermo field dynamics conception \cite{c31,c32} using the nonequilibrium
statistical operator method \cite{c21,c33,c34}. In this approach, similarly
to \cite{c20,c21}, the decisive role is that a set of the observed 
quantities is included in the description of the nonequilibrium process. For 
these quantities one finds generalized transport equations which should
agree with nonequilibrium thermodynamics at controlling the local
conservation laws for the particles-number density, momentum and energy.  It 
gives substantial advantages over the nonequilibrium Green function technique 
\cite{c14,c15,c16}, which quite well describes excitation spectra, but
practically does not describe nonequilibrium thermodynamics, and has
problems with the local conservation laws control and the generalized 
transport coefficients calculation.

In this paper we consider the kinetics and  hydrodynamics of highly
nonequilibrium and strongly coupled quantum systems using the nonequilibrium 
thermo field dynamics on the basis of the D.N.Zubarev nonequilibrium 
statistical operator method \cite{c27}. Within this method we consider two 
different levels of a description of the kinetics and hydrodynamics of dense 
quantum nuclear systems: strongly coupled states and a quark-gluon plasma. 
The nonequilibrium thermo field dynamics on the basis of the nonequilibrium 
statistical operator method constitutes sections 2 to 7. A nonequilibrium 
thermo vacuum state vector is obtained here in view of equations for the 
generalized hydrodynamics of dense quantum systems. Transport equations of a 
consistent description of the kinetics and hydrodynamics in thermo field 
representation are obtained in section 8.  We mean that these equations are 
applied to dense quantum systems where strong coupled states can appear. 
This item implies, as one of the approaches, to investigate a nonequilibrium 
nuclear matter \cite{c8,c9}. Another level of the description (rather a 
microscopic one) of nonequilibrium properties of dense quantum systems is 
considered in section 9. A consistent description of the kinetics and 
hydrodynamics is obtained here for a quark-gluon plasma.

\section{Thermo field dynamics formalism. Superoperators and state vectors
in the Liouville thermo field space}

In this section we reformulate the nonequilibrium statistical mechanics of
quantum systems using the thermo field dynamics formalism \cite{c31,c32}.

Let us consider a quantum system of $N$ interacting bosons or fermions. The
Hamiltonian of this system is expressed via creation $\av_l$ and
annihilation $\ai_l$ operators of the corresponding statistics:
%
%
\be
H=H(a^+,a).\label{e2.1}
\ee
Operators $\av_l$, $\ai_l$ satisfy the commutation relations:
%
%
\be
[\ai_l,\av_j]_\sigma=\delta_{lj},\quad
[\ai_l,\ai_j]_\sigma=[\av_l,\av_j]_\sigma=0,\label{e2.2}
\ee
where $[A,B]_\sigma=AB-\sigma BA$, $\sigma=+1$ for bosons and $\sigma=-1$
for fermions.

The nonequilibrium state of such a system is completely described by the 
non\-equ\-i\-li\-bri\-um statistical operator $\vrho(t)$. This operator 
satisfies the quantum Liouville equation
%
%
\be
\ddt\vrho(t)-\frac{1}{\im\hbar}[H,\vrho(t)]=0.\label{e2.3}
\ee
The nonequilibrium statistical operator $\vrho(t)$ allows us to calculate
the average values of operators $A$
%
%
\be
\la A\ra^t=\Sp\lp A\vrho(t)\rp,\label{e2.4}
\ee
which can be observable quantities describing the nonequilibrium state
of the system (for example, a hydrodynamic state is described by the average
values of operators of particle number, momentum and energy densities).

The main idea of thermo field dynamics \cite{c31,c32} and its nonequilibrium
formulation \cite{c35,c36,c37} consists in doing the calculation of average 
values \refp{e2.4} with the help of the so-called nonequilibrium thermo
vacuum state vector:
%
%
\be
\la A\ra^t=\la\la 1|A\vrho(t)\ra\ra=\la\la 1|\hat{A}|\vrho(t)\ra\ra,
\label{e2.5}
\ee
where $\hat{A}$ is a superoperator which acts on state $|\vrho(t)\ra\ra$.
Nonequilibrium thermo vacuum state vector $|\vrho(t)\ra\ra$ satisfies the
Schr\"odinger equation. Starting from equation \refp{e2.3}, we obtain the
relation
\[
\ddt|\vrho(t)\ra\ra-
\left.\left|\frac{1}{\im\hbar}[H,\vrho(t)]\right\ra\!\!\right\ra=0,
\]
or, opening commutator,
%
%
\be
\ddt|\vrho(t)\ra\ra-\frac{1}{\im\hbar}\bar{H}|\vrho(t)\ra\ra=0.\label{e2.6}
\ee
Here the ``total'' Hamiltonian $\bar{H}$ reads:
%
%
\be
\bar{H}=\hat{H}-\tilde{H},\label{e2.7}
\ee
and it is known that $\la\la 1|\bar{H}=0$; $\hat{H}=H(\hav,\hat{a})$,
$\tilde{H}=H^*(\tav,\tilde{a})$ are superoperators which consist of
creation and annihilation superoperators without and with a tilde, and which
represent the thermal Liouville space~\cite{c38,c39}. Superoperators
$\hat{H}$ and $\tilde{H}$ are defined in accordance with the relations:
%
%
\begin{equation}
\begin{array}{l}
|H\vrho(t)\ra\ra=\hat{H}|\vrho(t)\ra\ra,\\
|\vrho(t)H\ra\ra=\tilde{H}|\vrho(t)\ra\ra.\\
\end{array}
\label{e2.8}
\end{equation}
Hence, it appears that at going from the quantum Liouville equation
\refp{e2.3} for nonequilibrium statistical operator $\vrho(t)$ to the
Schr\"odinger equation \refp{e2.6} for non\-equi\-lib\-ri\-um thermo vacuum
state vector $|\vrho(t)\ra\ra$, according to \refp{e2.5}, the number of
creation and annihilation operators is doubled. Superoperators
$\hav_l$, $\hai_j$, $\tav_l$, $\tai_j$ satisfy the same commutation
relations as for operators $\av_l$, $\ai_j$ \refp{e2.2}:
%
%
\be
[\hat{a}_l,\hat{a}^+_j]_\sigma=[\tilde{a}_l,\tilde{a}^+_j]_\sigma=
\delta_{lj},\qquad
[\hat{a}_l,\tilde{a}_j]_\sigma=[\hat{a}^+_l,\tilde{a}^+_j]_\sigma=0,
\label{e2.9}
\ee
\[
[\hat{a}_l,\hat{a}_j]_\sigma=[\hat{a}^+_l,\hat{a}^+_j]_\sigma=0,\qquad\;\,
[\tilde{a}_l,\tilde{a}_j]_\sigma=[\tilde{a}^+_l,\tilde{a}^+_j]_\sigma=0.
\]
Annihilation superoperators $\hai_l$, $\tai_l$ are defined in accordance with
their action on the vacuum state -- the supervacuum \cite{c38}
%
%
\be
\hat{a}_l|00\ra\ra=\tilde{a}_l|00\ra\ra=0,\label{e2.10}
\ee
where $|00\ra\ra=||0\ra\la 0|\ra\ra$ is a supervacuum; and it is 
known that $\hai_l|0\ra=\ai_l|0\ra=0$, and $\la 0|\tai_l=0$, i.e. a
supervacuum $|00\ra\ra$ is an orthogonalized state of two va\-cu\-um states
$\la 0|$ and $|0\ra$. Taking into account commutation relations \refp{e2.9},
\refp{e2.10}, one can introduce unit vectors
$| 1\ra\ra=|\sum_l|l\ra\la l|\ra\ra$ and
$\la\la 1|=\la\la\sum_l|l\ra\la l||$ in the following forms:
%
%
\be
\begin{array}{l}
|1\ra\ra=\exp\lc\suml_l\hav_l\tav_l\rc|00\ra\ra,\\
\la\la 1|=\la\la 00|\exp\lc\suml_l\tai_l\hai_l\rc.\\
\end{array}
\label{e2.11}
\ee
With the help of these expressions one can find relations between the 
action of superoperators $\hav_l$, $\hai_j$, $\tav_l$, $\tai_j$
%
%
\be
\begin{array}{rcrlcl}
\hai_l|1\ra\ra&=&\tav_l|1\ra\ra,\quad&\la\la 1|\hav_l&=&\la\la 1|\tai_l,\\
\hav_l|1\ra\ra&=&\sigma\tai_l|1\ra\ra,\quad&\la\la 1|\hai_l&=&\la\la
1|\tav_l\sigma.
\end{array}
\label{e2.12}
\ee
In such a way, in the thermal field dynamics formalism \cite{c31,c32} the
number of operators is doubled by introducing both without tilde and
tildian operators $A(\hat{a}^+,\hat{a})$, $\tilde{A}(\tilde{a}^+,\tilde{a})$
for which the following properties take place:
%
%
\be
\begin{array}{rcl}
\widetilde{A_1A_2}&=&\tilde{A}_1\tilde{A}_2,\quad\tilde{\tilde{A}}=A,\\
\widetilde{c_1A_1+c_2A_2}&=&c_1^*\tilde{A}_1+c_2^*\tilde{A}_2,\\
|A\ra\ra&=&\hat{A}|1\ra\ra,\\
|A_1A_2\ra\ra&=&\hat{A}_1|A_2\ra\ra.\\
\end{array}
\label{e2.13}
\ee
Here $^*$ denotes a complex conjugation. Some detailed description of the 
properties of superoperators $\hav_l$, $\hai_j$, $\tav_l$, $\tai_j$, as well
as a thermal Liouville space is given in papers \cite{c31,c32,c38,c39}.

The nonequilibrium thermo vacuum state vector is normalized
%
%
\be
\la\la 1|\vrho(t)\ra\ra=\la\la 1|\hat{\vrho}(t)|1\ra\ra=1,\label{e2.14}
\ee
where $\hat{\vrho}(t)$ is a nonequilibrium statistical superoperator. It
depends on superoperators $\av_l$, $\ai_l$
%
%
\be
\hat{\vrho}(t)\equiv\vrho\lp\hav,\hat{a};t\rp,\label{e2.15}
\ee
and, it is known that the corresponding tildian superoperator
$\tilde{\vrho}(t)\equiv\vrho^+(\tav,\tilde{a};t)$ depends on superoperators
$\tav_l$, $\tai_l$.

\section{Nonequilibrium statistical operator in thermo field space}

To solve the Schr\"odinger equation \refp{e2.6} a boundary condition
should be given. Following the nonequilibrium statistical operator method
\cite{c21,c27,c33,c34}, let us find a solution to this equation in a
form, which depends on time via some set of observable quantities only. It
means that this set is sufficient for the description of a nonequilibrium
state of a system and does not depend on the initial moment of time. The
solution to the Schr\"odinger equation, which satisfies the following
boundary condition
%
%
\be
|\vrho(t)\ra\ra_{t=t_0}=|\vrhoq(t_0)\ra\ra,\label{e3.1}
\ee
reads:
%
%
\be
|\vrho(t)\ra\ra=\exp\lc(t-t_0)\frac{1}{\im\hbar}\bar{H}\rc
	|\vrhoq(t_0)\ra\ra.\label{e3.2}
\ee
We will consider times $t\gg t_0$, when the details of the initial state 
become inessential. To avoid the dependence on $t_0$, let us average the 
solution \refp{e3.2} on the initial time moment in the range between $t_0$ 
and $t$ and make the limiting transition $t_0-t\to-\infty$. We will obtain 
\cite{c27}:
%
%
\be
|\vrho(t)\ra\ra=\veps\intl_{-\infty}^0\d t'\;\e^{\veps t'}
	\e^{-\frac{1}{\im\hbar}\bar{H}t}|\vrhoq(t+t')\ra\ra,\label{e3.3}
\ee
where $\veps\to+0$ after the thermodynamic limiting transition. Solution 
\refp{e3.3}, as it can be shown by its direct dif\-fe\-ren\-ti\-a\-tion
with respect to time $t$, satisfies the Schr\"odin\-ger equation with a
small source in the right-hand side:
%
%
\be
\lp\ddt-\frac{1}{\im\hbar}\bar{H}\rp|\vrho(t)\ra\ra=
	-\veps\Big(|\vrho(t)\ra\ra-|\vrhoq(t)\ra\ra\Big).\label{e3.4}
\ee
This source selects retarded solutions which correspond to a shortened
description of the nonequilibrium state of a system, $|\vrhoq(t)\ra\ra$ is a
thermo vacuum quasiequilibrium state vector
%
%
\be
|\vrhoq(t)\ra\ra=\hvrhoq(t)|1\ra\ra.\label{e3.5}
\ee
Similarly to \refp{e2.14}, it is normalized by the rule
%
%
\be
\la\la 1|\vrhoq(t)\ra\ra=\la\la 1|\hvrhoq(t)|1\ra\ra=1,\label{e3.6}
\ee
where $\hvrhoq(t)$ is a quasiequilibrium statistical superope\-ra\-tor. The 
quasiequilibrium ther\-mo vacuum state vector of a system is introduced in 
the following way. Let $\la p_n\ra^t=\la\la 1|\hat{p}_n|\vrho(t)\ra\ra$ be 
a set of observable quantities which describe the nonequilibrium state of 
a system. $p_n$ are operators which consist of the creation and annihilation
operators defined in \refp{e2.2}. Quasiequilibrium statistical operator
$\vrhoq(t)$ is defined from the condition of informational entropy $S_{\rm
inf}$ extremum (maximum) at additional conditions of prescribing the average
values $\la p_n\ra^t$ and conservation of normalization condition
\refp{e3.6} \cite{c21,c34}:
%
%
\be
\begin{array}{rcl}
\ds\vrhoq(t)&=&\ds\exp\lc-\Phi(t)-\sum_nF_n^*(t)p_n\rc,\\\\
\ds\Phi(t)&=&\ds\ln\Sp\exp\lc-\sum_nF_n^*(t)p_n\rc,\\
\end{array}
\label{e3.7}
\ee
where $\Phi(t)$ is the Massieu-Planck functional. A summation on $n$ can
designate a sum with respect to the wave-vector $\bmv{k}$, the kind of 
particles and a line of other quantum numbers, a spin for example. 
Parameters $F_n(t)$ are defined from the conditions of self-consistency:
%
%
\be
\la p_n\ra^t=\ds\la p_n\ra^t_{\rm q},\qquad
	\la\ldots\ra^t_{\rm q}=\ds\Sp\Big(\ldots\vrhoq(t)\Big).
	\label{e3.8}
\ee
According to \refp{e2.5}, let us write these conditions of self-consistency 
in the following form:
%
%
\be
\la\la 1|\hat{p}_n|\vrho(t)\ra\ra=\la\la 1|\hat{p}_n|\vrhoq(t)\ra\ra.
\label{e3.9}
\ee
Taking into account behaviours \refp{e2.13}, we have:
%
%
\be
|\vrhoq(t)\ra\ra=\hvrhoq(t)|1\ra\ra=\tilde{\vrho}^+_{\rm q}(t)|1\ra\ra,
\label{e3.10}
\ee
where
%
%
\be
\begin{array}{lcl}
\ds\hvrhoq(t)&=&\ds\exp\lc-\Phi(t)-\sum_nF_n^*(t)\hat{p}_n\rc,\\\\
\ds\tilde{\vrho}_{\rm q}^+(t)&=&\ds\exp\lc-\Phi(t)-\sum_nF_n(t)
	\tilde{p}_n\rc\\
\end{array}
\label{e3.11}
\ee
are quasiequilibrium statistical superoperators which contain 
superoperators $\hat{p}_n$ and $\tilde{p}_n$, correspondingly:
%
%
\be
\begin{array}{lcl}
\hat{p}_n&=&p_n(\hav,\hat{a}),\\
\tilde{p}_n&=&p_n^*(\tav,\tilde{a}).\\
\end{array}
\label{e3.12}
\ee
If self-consistency condition \refp{e3.9} realizes, we shall have the
following relations (at fixed corresponding parameters):
%
%
\be
\begin{array}{l}
\ds\frac{\delta\Phi(t)}{\delta F_n^*(t)}=\la\la 1|\hat{p}_n|\vrhoq(t)\ra\ra
	=\la\la 1|\hat{p}_n|\vrho(t)\ra\ra,\\\\
\ds\frac{\delta\Phi(t)}{\delta F_n(t)}=\la\la 1|\tilde{p}_n|\vrhoq(t)\ra\ra
	=\la\la 1|\tilde{p}_n|\vrho(t)\ra\ra.\\
\end{array}
\label{e3.13}
\ee
Relations \refp{e3.13} show that parameters $F_n^*(t)$, $F_n(t)$ are
conjugated to averages $\la\la 1|\hat{p}_n|\vrho(t)\ra\ra$ and
$\la\la 1|\tilde{p}_n|\vrho(t)\ra\ra$, correspondingly. On the other hand,
with the help of $|\vrhoq(t)\ra\ra$ and self-consistency conditions
\refp{e3.9} we can define the entropy of the system state:
%
%
\be
S(t)=-\la\la 1|\big(\ln\vrhoq(t)\big)\vrhoq(t)\ra\ra=
\Phi(t)+\sum_nF_n^*(t)\la\la 1|\hat{p}_n|\vrho(t)\ra\ra.\label{e3.14}
\ee
The physical meaning of parameters $F_n^*(t)$  can be obtained now on the
basis of the previous relation:
%
%
\be
F_n^*(t)=\frac{\delta S(t)}{\delta\la\la 1|\hat{p}_n|\vrho(t)}.\label{e3.15}
\ee

\section{Projection operators in thermo field space}

Now the auxiliary quasiequilibrium thermo vacuum state vector
$|\vrhoq(t)\ra\ra$ is defined. Let us represent solution \refp{e3.3} of
the Schr\"odinger equation \refp{e2.6} in a form which is more convenient
for the construction of transport equations for averages
$\la\la 1|\hat{p}_n|\vrho(t)\ra\ra$. We shall start from the Schr\"odinger
equation with a small source \refp{e3.4}. Let us rebuild this equation by
introducing $\vt|\vrho(t)\ra\ra=|\vrho(t)\ra\ra-|\vrhoq(t)\ra\ra$:
%
%
\be
\lp\ddt-\frac{1}{\im\hbar}\bar{H}+\veps\rp\vt|\vrho(t)\ra\ra=
-\lp\ddt-\frac{1}{\im\hbar}\bar{H}\rp|\vrhoq(t)\ra\ra.\label{e4.1}
\ee
The calculation of time derivation of $|\vrhoq(t)\ra\ra$ in the right-hand
side of equation \refp{e4.1} is equivalent to the introduction of the
Kawasaki-Gunton projection operator $\SP_{\rm q}(t)$ \cite{c27} in thermo
field representation:
%
%
\be
\ddt|\vrhoq(t)\ra\ra=\SP_{\rm q}(t)\frac{1}{\im\hbar}\bar{H}|\vrho(t)\ra\ra,
\label{e4.2}
\ee
%
%
\bea
\lefteqn{\ds\SP_{\rm q}(t)\big(|\ldots\ra\ra\big)=|\vrhoq(t)\ra\ra+{}}
	\label{e4.3}\\
\lefteqn{\ds\sum_n\frac{\delta|\vrhoq(t)\ra\ra}
	{\delta\la\la 1|\hat{p}_n|\vrho(t)\ra\ra}
	\la\la 1|\hat{p}_n|\ldots\ra\ra-
	\sum_n\frac{\delta|\vrhoq(t)\ra\ra}
	{\delta\la\la 1|\hat{p}_n|\vrho(t)\ra\ra}
	\la\la 1|\hat{p}_n|\ldots\ra\ra\la\la 1|\ldots\ra\ra.}\nonumber
\eea
Projection operator $\SP_{\rm q}(t)$ acts on state vectors $|\ldots\ra\ra$
only and has all the operator properties:
\[
\begin{array}{lcl}
\ds\SP_{\rm q}(t)|\vrho(t')\ra\ra&=&|\vrhoq(t)\ra\ra,\phantom{\intl_0}\\
\ds\SP_{\rm q}(t)|\vrhoq(t')\ra\ra&=&|\vrhoq(t)\ra\ra,\phantom{\intl_0}\\
\ds\SP_{\rm q}(t)\SP_{\rm q}(t')&=&\SP_{\rm q}(t).\phantom{\intl_0}\\
\end{array}
\]
Taking into account condition
$\SP_{\rm q}(t)\frac{1}{\im\hbar}\bar{H}\vt|\vrho(t)\ra\ra=0$, one may
rewrite equation \refp{e4.1}, after simple reductions, in a form:
%
%
\be
\lp\ddt-\Big(1-\SP_{\rm q}(t)\Big)\frac{1}{\im\hbar}\bar{H}+\veps\rp
	\vt|\vrho(t)\ra\ra=
	\ds\Big(1-\SP_{\rm q}(t)\Big)\frac{1}{\im\hbar}\bar{H}
	|\vrhoq(t)\ra\ra.\label{e4.4}
\ee
The formal solution to this equation reads:
\[
\vt|\vrho(t)\ra\ra=\intl_{-\infty}^t\d t'\;\e^{\veps(t'-t)}T(t,t')
	\Big(1-\SP_{\rm q}(t')\Big)\frac{1}{\im\hbar}\bar{H}
	|\vrhoq(t')\ra\ra,
\]
or
%
%
\be
|\vrho(t)\ra\ra=|\vrhoq(t)\ra\ra+\intl_{-\infty}^t\d t'\;\e^{\veps(t'-t)}
	T(t,t')\Big(1-\SP_{\rm q}(t')\Big)\frac{1}{\im\hbar}\bar{H}
	|\vrhoq(t')\ra\ra,\label{e4.5}
\ee
where
%
%
\be
T(t,t')=\exp_+\lc\int_{t'}^t\d t'\;
	\Big(1-\SP_{\rm q}(t')\Big)\frac{1}{\im\hbar}\bar{H}\rc
	\label{e4.6}
\ee
is an evolution operator with projection consideration, and $\exp_+$ is an 
ordered exponent. Then, let us consider expression
$\Big(1-\SP_{\rm q}(t')\Big)\frac{1}{\im\hbar}\bar{H}|\vrhoq(t)\ra\ra$ in
the right-hand side of \refp{e4.1}. The action of
$\frac{1}{\im\hbar}\bar{H}$ and $\Big(1-\SP_{\rm q}(t')\Big)$ on
$|\vrhoq(t)\ra\ra$ can be represented in the form:
%
%
\be
\Big(1-\SP_{\rm q}(t')\Big)\frac{1}{\im\hbar}\bar{H}|\vrhoq(t)\ra\ra=
	\sum_nF_n^*(t)\left.\left|
	\mbox{$\intl_0^1$}\d\tau\;\vrhoq^\tau(t)
	\Big(1-\SP(t')\Big)\dot{p}_n\vrhoq^{1-\tau}(t)\Ra\!\!\Ra\!,
	\label{e4.7}
\ee
where $\dot{p}_n$ and $\SP(t)$ read:
%
%
\bea
\dot{p}_n&=&-\frac{1}{\im\hbar}[H,p_n],\label{e4.8}\\
\SP(t)p&=&\la\la 1|\hat{p}|\vrhoq(t)\ra\ra+
	\sum_n\frac{\delta\la\la 1|\hat{p}|\vrhoq(t)\ra\ra}
	{\delta\la\la 1|\hat{p}_n|\vrho(t)\ra\ra}
	\Big(p_n-\la\la 1|\hat{p}_n|\vrho(t)\ra\ra\Big).\label{e4.9}
\eea
Here $\SP(t)$ is a generalized Mori projection operator in thermo field
representation. It acts on operators and has the following properties:
%
%
\be
\begin{array}{lcl}
\SP(t)p_n&=&p_n,\\
\SP(t)\SP(t')&=&\SP(t).\\
\end{array}
\label{e4.10}
\ee
Let us substitute now \refp{e4.7} into \refp{e4.5} and, as a result, we will
obtain an expression for the nonequilibrium thermo vacuum state of a system:
%
%
\bea
\lefteqn{\ds\ds|\vrho(t)\ra\ra=|\vrhoq(t)\ra\ra+
\sum_n\intl_{-\infty}^t\d t'\e^{\veps(t'-t)}T(t,t')
	\left.\left|\mbox{$\intl_0^1$}\d\tau\vrhoq^\tau(t')J_n(t')
	\vrhoq^{1-\tau}(t')\Ra\!\!\Ra\! F_n^*(t').}\nonumber\\
	\label{e4.11}
\eea
Here
%
%
\be
J_n(t)=\Big(1-\SP(t)\Big)\dot{p}_n\label{e4.12}
\ee
\nopagebreak
are generalized flows.
\pagebreak

Let us obtain now transport equations for averages
$\la\la 1|\hat{p}_n|\vrho(t)\ra\ra$ in thermo field representation with the
help of nonequilibrium thermo vacuum state vector $|\vrho(t)\ra\ra$
\refp{e4.11}. To achieve this we will use the equality
%
%
\be
\ddt\la\la 1|\hat{p}_n|\vrho(t)\ra\ra=
	\la\la 1|\dot{\hat{p}}_n|\vrho(t)\ra\ra=
	\la\la 1|\dot{\hat{p}}_n|\vrhoq(t)\ra\ra+
	\la\la J_n(t)|\vrho(t)\ra\ra.\label{e4.13}
\ee
By making use of $|\vrho(t)\ra\ra$ in \refp{e4.11} in averaging the last 
term, we obtain transport equations for $\la\la 1|\hat{p}_n|\vrho(t)\ra\ra$:
%
%
\bea
\lefteqn{\ds\ddt\la\la 1|\hat{p}_n|\vrho(t)\ra\ra=
	\la\la 1|\dot{\hat{p}}_n|\vrhoq(t)\ra\ra+{}}\label{e4.14}\\
\lefteqn{\ds\sum_{n'}\intl_{-\infty}^t\d t'\;\e^{\veps(t'-t)}
	\La\!\!\La J_n(t)T(t,t')\left|\mbox{$\intl_0^1$}\d\tau\;
	\vrhoq^\tau(t')J_{n'}(t')\vrhoq^{1-\tau}(t')\Ra\!\!\Ra\right.
	F_{n'}^*(t'),}\nonumber
\eea
where $\dot{\hat{p}}_n=-\frac{1}{\im\hbar}\bar{H}\hat{p}_n$.
Relations \refp{e4.14} are treated as a general form of transport equations
for average values of a shortened description. These equations can be
applied to completely actual problems.

In the case of weakly nonequilibrium processes, the generalized transport
equations \refp{e4.14} are reduced appreciably. We shall consider this case
in the next section.

\section{Transport equations in linear approximation}

Let us consider the nonequilibrium state of a quantum field system near
equilibrium. In this connection let us suppose that average values $\la\la
1|\hat{p}_n|\vrho(t)\ra\ra$ of variables for a shortened description
and their conjugated parameters $F^*_n(t)$ differ in their equilibrium
meanings slightly. In such a case, one can expand the quasiequilibrium
thermo vacuum state vector $|\vrhoq(t_0)\ra\ra$ \refp{e3.5}, \refp{e3.11}
into a series on deviations of parameters $F^*_n(t)$ from their equilibrium
values $F_n(0)$ and restrict a linear approximation
only:
%
%
\be
|\vrhoq(t)\ra\ra=|\vrho_0(F_n(0))\ra\ra-
	\sum_n\delta F_n^*(t)\left.\left|\mbox{$\intl_0^1$}\d\tau\;
	\vrho_0^\tau(F_n(0))p_n\vrho_0^{1-\tau}(F_n(0))\Ra\!\!\Ra,
	\label{e5.1}
\ee
where $|\vrho_0(F_n(0))\ra\ra$ is an equilibrium thermo vacuum sta\-te
vector which depends on equilibrium values of $F_n(0)$ parameters (local
inversed temperature $\beta$ and chemical potential $\mu$),
$\delta F_n^*(t)=F^*_n(t)-F^*_n(0)$. With the help of self-consistency
conditions \refp{e3.9} and taking into account \refp{e5.1}, let us define
parameters $\delta F^*_n(t)$:
%
%
\be
\delta F_n^*(t)=-\sum_m((p|p))^{-1}_{nm}
	\la\la 1|\delta\hat{p}_m|\vrho(t)\ra\ra,\label{e5.2}
\ee
where $\delta\hat{p}_m=\hat{p}_m-\la\la 1|\hat{p}_m|\vrho_0(F_n(0))\ra\ra$,
$((p|p))_{mn}^{-1}$ are elements of the inverse matrix of $((p|p))$.
Elements of the matrix $((p|p))$ are equilibrium correlation functions in
thermo field representation
%
%
\be
((p_n|p_m))_0=\La\!\!\La p_n\left|\mbox{$\intl_0^1$}\d\tau\;
	\vrho_0^\tau p_m\vrho_0^{1-\tau}\right.\Ra\!\!\Ra.\label{e5.3}
\ee
Substitution of \refp{e5.3} into \refp{e5.1} for $|\vrhoq(t)\ra\ra$ results
in
%
%
{\small
\bea
\lefteqn{\ds|\vrhoq(t)\ra\ra=|\vrho_0(F(0))\ra\ra+
	\sum_{m,n}\la\la 1|\delta\hat{p}_m|\vrho(t)\ra\ra
	((p|p))^{-1}_{mn}\left.\left|\mbox{$\intl_0^1$}\d\tau\;
	\vrho_0^\tau(F(0))p_n\vrho_0^{1-\tau}(F(0))
	\Ra\!\!\Ra.}\nonumber\\
	\label{e5.4}
\eea}

\vspace*{-2ex}\noindent
To calculate nonequilibrium thermo vacuum state vector $|\vrho(t)\ra\ra$
\refp{e4.11} in linear approximation \refp{e5.4}, let us rewrite transport 
equations \refp{e4.14} in another form:
%
%
{\small
\bea
\lefteqn{\ds\ddt\la\la 1|\delta\hat{p}_n|\vrho(t)\ra\ra=
	\sum_m\Omega_{nm}\la\la 1|\delta\hat{p}_n|\vrho(t)\ra\ra-
	\sum_m\mbox{$\intl_{-\infty}^t$}
	\d t'\;\e^{\veps(t'-t)}\vphi_{nm}(t,t')
	\la\la 1|\delta\hat{p}_n|\vrho(t')\ra\ra,}\nonumber\\
	\label{e5.5}
\eea}

\vspace*{-2ex}\noindent
where
%
%
\be
\Omega_{nm}=\sum_l\La\!\!\La\dot{p}_n\left|\mbox{$\intl_0^1$}\d\tau\;
	\vrho_0^\tau p_l\vrho_0^{1-\tau}\right.\Ra\!\!\Ra
	((p|p))^{-1}_{lm}\label{e5.6}
\ee
are equilibrium quantum correlation functions, and
%
%
\be
\vphi_{nm}(t,t')=\sum_l\La\!\!\La J_n\left|T_0(t,t')\left|
	\mbox{$\intl_0^1$}\d\tau\;
	\vrho_0^\tau J_l\vrho_0^{1-\tau}\right.\right.\Ra\!\!\Ra
	((p|p))^{-1}_{lm}\label{e5.7}
\ee
are transport cores in thermo field representation, which de\-scribe
dissipative processes in the weakly nonequilibrium state of a quantum 
system, $J_n$ is defined similarly to \refp{e4.12}:
\bean
J_n&=&(1-\SP_0)\dot{p}_n;\\
T_0(t)&=&\exp\lc t(1-\SP_0)\frac{\im}{\hbar}\bar{H}\rc
\eean
is the time evolution operator with taking into account the projecting where
$\SP_0$ is the Mori projection operator with the following structure:
%
%
\be
\SP_0 A=\la\la 1|\hat{A}|\vrho_0\ra\ra+
	\sum_{m,n}((A|p_n))((p|p))^{-1}_{nm}p_m.\label{e5.8}
\ee
Projection operator $\SP_0$ satisfies conditions \refp{e4.10}. As it can be
shown \cite{c27}, for transport equations \refp{e5.5} correspond to
equations for quantum time correlation functions in thermo field
representation:
%
%
\bea
\Phi_{nm}(t)&=&\la\la p_n(t)|p_m(0)\vrho_0\ra\ra,\label{e5.9}\\
\ddt\Phi_{nm}(t)&=&\sum_l\Omega_{nl}\Phi_{lm}(t)-
	\sum_l\intl_{-\infty}^t\d t'\;\e^{\veps(t'-t)}\vphi_{nl}(t,t')
	\Phi_{lm}(t'),\label{e5.10}
\eea
where the time evolution of superoperators $\hat{p}_n(t)$ in the Heisenberg
representation reads:
%
%
\be
\label{e5.11}\hat{p}_n(t)=
\exp\lc-\frac{1}{\im\hbar}\bar{H}t\rc
	\hat{p}_n\exp\lc\frac{1}{\im\hbar}\bar{H}t\rc.
\ee
As it is indicated by \refp{e5.5}, \refp{e5.10}, transport equations for
values $\la\la 1|\hat{p}_n|\vrho(t)\ra\ra$ and the corresponding
equations for time correlation functions $\Phi_{nm}(t)$ in linear
approximation \refp{e5.4} are closed.

If spectral magnitudes $\Phi_{nm}(\omega)$ for quantum time correlation
functions $\Phi_{nm}(t)$ are defined as follows:
%
%
\be
\Phi_{nm}(\omega)=\frac{1}{2\pi}\intl_{-\infty}^{\infty}\d t\;
	\e^{\im\omega t}\Phi_{nm}(t),\label{e5.12}
\ee
then, according to the definition of Green functions \cite{c21,c33} and the
condition that $\Phi_{nm}(\omega)$ is a real function ($\omega$ is a real
number), one finds imaginary parts of spectral magnitudes of the 
corresponding retarded, advanced and causal Green functions
%
%
\bea
\Imm G^{\rm r,a}_{nm}(\omega)&=&\mp\frac{1}{2\hbar}
	\lp\e^{\beta\hbar\omega}-\sigma\rp\Phi_{nm}(\omega),\label{e5.13}\\
\Imm G^{\rm c}_{nm}(\omega)&=&-\;\frac{1}{2}\;
	\lp\e^{\beta\hbar\omega}+\sigma\rp\Phi_{nm}(\omega),\label{e5.14}
\eea
where functions $G^{\rm r}_{nm}(t,t')$, $G^{\rm a}_{nm}(t,t')$,
$G^{\rm c}_{nm}(t,t')$ in time representation constitute, correspondingly, 
retarded, advanced and causal Green functions in thermo field representation:
%
%
\bea
G^{\rm r,a}_{nm}(t,t')&=&\pm\frac{1}{\im\hbar}\theta(t-t')
	\la\la 1|[p_n(t),p_m(t')]_\sigma|\vrho_0\ra\ra={}\label{e5.15}\\
&&\pm\frac{1}{\im\hbar}\theta(t-t')
	\la\la 1|[\hat{p}_n(t),\hat{p}_m(t')]_\sigma|\vrho_0\ra\ra={}
	\nonumber\\
&&\pm\frac{1}{\im\hbar}\theta(t-t')
	\la\la 1|\hat{p}_n(t)\hat{p}_m(t')-
	\sigma\hat{p}_m(t')\hat{p}_n(t)|\vrho_0\ra\ra,\nonumber
\eea
%
%
%
%
%
\bea
G^{\rm c}_{nm}(t,t')&=&\frac{1}{\im\hbar}\la\la 1|Tp_n(t)p_m(t')\vrho_0\ra\ra
	\label{e5.16}\\
&=&\frac{1}{\im\hbar}\theta(t-t')
	\la\la 1|\hat{p}_n(t)\hat{p}_m(t')|\vrho_0\ra\ra+
	\frac{1}{\im\hbar}\theta(t-t')
	\la\la 1|\tilde{p}^+_n(t)\hat{p}_m(t')|\vrho_0\ra\ra,\nonumber
\eea
where $\theta(t)=\lc
\begin{array}{ll}
1,&t>0,\\
0,&t<0,\\
\end{array}\right.$ is a unit step function. As it is seen from \refp{e5.15}
and \refp{e5.16}, retarded (advanced) and causal Green functions are
defined via time correlation functions
$\la\la 1|\hat{p}_n(t)\hat{p}_m(t')|\vrho_0\ra\ra$,
$\la\la 1|\tilde{p}_n^+(t)\hat{p}_m^{\phantom{+}}(t')|\vrho_0\ra\ra$, which
satisfy conditions \refp{e5.10}. For completeness, let us write dispersion
relations which connect the imaginary parts $\Imm G^{\rm r,a}_{nm}(\omega)$,
$\Imm G^{\rm c}_{nm}(\omega)$ \refp{e5.13}, \refp{e5.14} and the real parts
$\Ree G^{\rm r,a}_{nm}(\omega)$, $\Ree G^{\rm c}_{nm}(\omega)$ of spectral
functions of the corresponding Green functions \cite{c21,c33}:
%
%
\bea
\lefteqn{\ds\Ree G^{\rm r,a}_{nm}(\omega)=\pm\frac{1}{\pi}\SP
	\intl_{-\infty}^\infty\d\omega'\;
	\frac{\Imm G^{\rm r,a}(\omega')}{\omega'-\omega},}\label{e5.17}\\
\lefteqn{\ds\Ree G^{\rm c}_{nm}(\omega)=\phantom{\pm}\frac{1}{\pi}\SP
	\intl_{-\infty}^\infty\d\omega'\;
	\frac{\Imm G^{\rm c}(\omega')}{\omega'-\omega}
	\frac{\e^{\beta\hbar\omega'}-\sigma}{\e^{\beta\hbar\omega'}+\sigma}.}
	\nonumber
\eea
It should be pointed out that Green functions in thermo field
representation, which are calculated with the help of equilibrium thermo
vacuum state vector $|\vrho_0(\beta)\ra\ra$, were investigated in many 
papers \cite{c32,c38,c40,c41}. Specifically, the diagram technique for
their calculation, which generalizes the Feynman method, was developed in
\cite{c42}.

An application of the general structure of nonequilibrium thermo field 
dynamics on the basis of the nonequilibrium statistical operator method 
\cite{c21,c27} will be considered in sections 7 and 8. We will obtain 
equations of hydrodynamics and transport equations of a consistent 
description of the kinetics and hydrodynamics for dense quantum systems in 
thermo field representation. While investigating such systems, one of 
important problems is the calculation of transport cores (or transport 
coefficients) both for weakly and strongly nonequilibrium systems.

The problem is that quasiequilibrium thermo vacuum
state vector $|\vrhoq(t)\ra\ra$, \refp{e3.10} or \refp{e5.4}, in each case 
is not a vacuum state for superoperators $\hat a$, $\hav$, $\tilde a$, 
$\tav$. The subject of this question consists in the construction of 
dynamical reflection of superoperators $\hat a$, $\hav$, $\tilde a$, $\tav$ 
into superoperators for ``quasiparticles'' for which the quasiequilibrium
thermo vacuum state is a vacuum state. One method of constructing such
superoperators is considered in the next section.

\section{Creation and annihilation superoperators of 
``quasiparticles'' for the quasiequilibrium thermo vacuum state}

In our further consideration we assume that Hamiltonian $H$ \refp{e2.1} 
of a system can be represented in the form:
%
%
\be
H=H_0+H_{\rm int},\label{e6.1}
\ee
where $H_{\rm int}$ contains a small parameter. This small parameter can be
used for the construction of perturbation theory series. At the same time 
$H_0$ is a non-perturbed part of the Hamiltonian \refp{e6.1}. It depends on 
creation and annihilation operators $\av_l$, $\ai_l$ bilinearly. 
According to the nonequilibrium thermo field dynamics formalism, 
Hamiltonian $\bar{H}$ of a system reads:
%
%
\be
\arraycolsep=0pt
\begin{array}{lclcl}
\bar{H}\;&=&\;\bar{H}_0&+&\bar{H}_{\rm int},\\
\bar{H}_0\;&=&\;\hat{H}_0&-&\tilde{H}_0,\\
\bar{H}_{\rm int}\;&=&\;\hat{H}_{\rm int}&-&\tilde{H}_{\rm int}.\\
\end{array}
\label{e6.2}
\ee
$\bar{H}_0$ depends on superoperators $\hav$, $\hat{a}$, $\tav$, $\tilde{a}$
bilinearly. In such a case, to construct the perturbation theory for
operators it is convenient to use the Heisenberg representation on the 
non-perturbed part of Hamiltonian $\bar{H}_0$:
%
%
\be
\bar{A}(t)=\exp\lc-\frac{1}{\im\hbar}\bar{H}_0t\rc\bar{A}
	\exp\lc\frac{1}{\im\hbar}\bar{H}_0t\rc.\label{e6.3}
\ee
It is known that $\bar{A}(t)$ satisfies the Heisenberg equation
%
%
\be
\ddt\bar{A}(t)=-\frac{1}{\im\hbar}\ls\bar{H}_{\rm int}(t),\bar{A}(t)\rs,
	\,\bar{H}_{\rm int}(t)=
	\exp\lc-\frac{1}{\im\hbar}\bar{H}_0t\rc\bar{H}
	\exp\lc\frac{1}{\im\hbar}\bar{H}_0t\rc\!.
	\label{e6.4}
\ee
Superoperators $\hav$, $\hat{a}$, $\tav$, $\tilde{a}$ in the Heisenberg
representation read:
%
%
\be
\arraycolsep=0pt
\begin{array}{lcllcl}
\hat{a}(t)&=&\;\e^{-\frac{1}{\im\hbar}\bar{H}_0t}\hat{a}
	\e^{\frac{1}{\im\hbar}\bar{H}_0t},&
	\qquad\hav(t)&=&\;\e^{-\frac{1}{\im\hbar}\bar{H}_0t}\hav
	\e^{\frac{1}{\im\hbar}\bar{H}_0t},\\
\tilde{a}(t)&=&\;\e^{-\frac{1}{\im\hbar}\bar{H}_0t}\tilde{a}
	\e^{\frac{1}{\im\hbar}\bar{H}_0t},&
	\qquad\tav(t)&=&\;\e^{-\frac{1}{\im\hbar}\bar{H}_0t}\tav
	\e^{\frac{1}{\im\hbar}\bar{H}_0t}.\\
\end{array}
\label{e6.5}
\ee
They satisfy the ``classical'' commutation relations:
%
%
\be
[\hat{a}(t),\hat{a}^+(t)]_\sigma=[\tilde{a}(t),\tilde{a}^+(t)]_\sigma=1,
\qquad
[\hat{a}(t),\tilde{a}(t)]_\sigma=[\hat{a}^+(t),\tilde{a}^+(t)]_\sigma=0,
\label{e6.6}
\ee
\[
[\hat{a}(t),\hat{a}(t)]_\sigma=[\hat{a}^+(t),\hat{a}^+(t)]_\sigma=0,
\qquad
[\tilde{a}(t),\tilde{a}(t)]_\sigma=[\tilde{a}^+(t),\tilde{a}^+(t)]_\sigma=0.
\]

Let us assume that the quasiequilibrium thermo vacuum state vector
describes the initial state of a quantum system by the non-perturbed part of
Hamiltonian $\bar{H}_0$. $|\vrhoq(t_0)\ra\ra$ is not a vacuum state for
annihilation superoperators $\hat{a}(t)$, $\tilde{a}(t)$, i.e.
%
%
\bea
\hat{a}(t)|\vrhoq(t_0)\ra\ra&=&\phantom{\sigma}f(t-t_0)\tav(t)
	|\vrhoq(t_0)\ra\ra,\nonumber\\
\tilde{a}(t)|\vrhoq(t_0)\ra\ra&=&\sigma f(t-t_0)\hav(t)
	|\vrhoq(t_0)\ra\ra,\label{e6.7}\\
\la\la 1|\hav(t)&=&\sigma\la\la 1|\tav(t),\nonumber
\eea
where function $f(t-t_0)$ will be defined below. Nevertheless, the linear
combination of superoperators $\hav$, $\hat{a}$, $\tav$, $\tilde{a}$ allows 
us to define new creation and annihilation superoperators $\hgv$,
$\hat{\gamma}(t)$ and $\tgv(t)$, $\tilde{\gamma}(t)$ \cite{c35,c36,c37}:
%
%
\bea
\hat{\gamma}(t)&=&Q^{1/2}(t-t_0)[\hat{a}(t)-f(t-t_0)\tav(t)],\label{e6.8}\\
\hgv(t)&=&Q^{1/2}(t-t_0)[\tav(t)-\sigma\hat{a}(t)].\label{e6.9}
\eea
So, taking into account \refp{e6.7}, the action of new operators on
state vectors reads:
%
%
\be
\arraycolsep=0pt
\begin{array}{lcl}
\hat{\gamma}(t)|\vrhoq(t_0)\ra\ra=0,&\quad&\la\la 1|\hgv(t)=0,\\
\tilde{\gamma}(t)|\vrhoq(t_0)\ra\ra=0,&\quad&\la\la 1|\tgv(t)=0.\\
\end{array}
\label{e6.10}
\ee
Superoperators $\hgv$, $\hat{\gamma}(t)$ and $\tgv(t)$, $\tilde{\gamma}(t)$
satisfy the ``canonical'' commutation relations:
%
%
\bea
\lefteqn{\ds[\hat{\gamma}(t),\hgv(t)]_\sigma=1,}\nonumber\\
\lefteqn{\ds[\tilde{\gamma}(t),\tgv(t)]_\sigma=1,\qquad
	[\hat{\gamma}(t),\tilde{\gamma}(t)]_\sigma=
	[\hgv(t),\tgv(t)]_\sigma=0.}\label{e6.11}
\eea
A connecting expression between multiplier $Q(t-t_0)$ and function
$f(t-t_0)$ may be found on the basis of relations \refp{e6.8}, \refp{e6.9}
and \refp{e6.11}:
%
%
\be
Q(t-t_0)=[1-\sigma f(t-t_0)]^{-1}.\label{e6.12}
\ee
But, to define $f(t-t_0)$ function let us use the second equality in
\refp{e6.7}. It aids to obtain the following:
%
%
\be
\la\la 1|\hat{a}(t)\tilde{a}(t)|\vrhoq(t_0)\ra\ra=\sigma f(t-t_0)
	\la\la 1|\hat{a}(t)\hav(t)|\vrhoq(t_0)\ra\ra.\label{e6.13}
\ee
And, using the third equality in \refp{e6.7} and \refp{e6.13} we arrive at
%
%
\be
\la\la 1|\tav(t)\tilde{a}(t)|\vrhoq(t_0)\ra\ra=f(t-t_0)
	\la\la 1|\hat{a}(t)\hav(t)|\vrhoq(t_0)\ra\ra.\label{e6.14}
\ee
As far as
%
%
\be
n(t-t_0)=\la\la 1|\tav(t)\tilde{a}(t)|\vrhoq(t_0)\ra\ra=
	\la\la 1|\hav(t)\hat{a}(t)|\vrhoq(t_0)\ra\ra\label{e6.15}
\ee
is the average particle number, then, using in the right-hand side of
\refp{e6.14} commutation relation
$\hat{a}(t)\hav(t)-\sigma\hav(t)\hat{a}(t)=1$, we obtain a linkage between
$n(t-t_0)$ and function $f(t-t_0)$, and vice versa:
%
%
\bea
n(t-t_0)&=&f(t-t_0)\big(1+\sigma n(t-t_0)\big),\nonumber\\
f(t-t_0)&=&n(t-t_0)/\big(1+\sigma n(t-t_0)\big).\label{e6.16}
\eea
Finally, if to substitute \refp{e6.16} into \refp{e6.12}, one finds a 
linkage of normalized multiplier $Q(t-t_0)$ and average particle number
$n(t-t_0)$:
%
%
\be
Q(t-t_0)=1+\sigma n(t-t_0).\label{e6.17}
\ee
Now, taking into account \refp{e6.16} and \refp{e6.17}, relations
\refp{e6.8}, \refp{e6.9} for superoperators $\hgv$, $\hat{\gamma}(t)$, and
$\tgv(t)$, $\tilde{\gamma}(t)$ read:
%
%
\bea
\hat{\gamma}(t)&=&\Big(1+\sigma n(t-t_0)\Big)^{\frac 12}
	\ls\hat{a}(t)-{\frac{n(t-t_0)}{1+\sigma n(t-t_0)}}
	\tav(t)\rs,\nonumber\\
\tgv(t)&=&\Big(1+\sigma n(t-t_0)\Big)^{\frac 12}\;
	[\tav(t)-\sigma\hat{a}(t)].\label{e6.18}
\eea
Inversed transformations for superoperators $\hav$, $\hat{a}(t)$ and
$\tav(t)$, $\tilde{a}(t)$ can be easily obtained from \refp{e6.18}:
%
%
\bea
\hat{a}(t)&=&\Big(1+\sigma n(t-t_0)\Big)^{\frac 12}
	\ls\hat{\gamma}(t)+{\frac{n(t-t_0)}{1+\sigma n(t-t_0)}}
	\tgv(t)\rs,\nonumber\\
\tav(t)&=&\Big(1+\sigma n(t-t_0)\Big)^{\frac 12}\;
	[\tgv(t)+\sigma\hat{\gamma}(t)].\label{e6.19}
\eea
$\hat{\gamma}(t)$, $\hgv(t)$, $\tilde{\gamma}(t)$, $\tgv(t)$ can be defined 
as annihilation and creation superoperators of quasiparticles for which 
quasiequilibrium thermo vacuum state $|\vrhoq(t)\ra\ra$ is a vacuum state. 
These superoperators are functions of thermodynamic parameters $F_n(t)$ 
which describe the quasiequilibrium state of a system. Relations 
\refp{e6.18}, \refp{e6.19} are a dynamical reflection between $\hat{a}(t)$, 
$\hav(t)$, $\tilde{a}(t)$, $\tav(t)$ and $\hat{\gamma}(t)$, $\hgv(t)$, 
$\tilde{\gamma}(t)$, $\tgv(t)$.  While calculating transport cores in 
transport equations \refp{e4.14} and \refp{e5.7}, or connected with these 
quantities Green functions, superoperators $\hat{a}(t)$, $\hav(t)$,
$\tilde{a}(t)$, $\tav(t)$ are multiplied according to the Wick theorem about 
normal products :$\hav(t)\hat{a}(t)$:, :$\tav(t)\tilde{a}(t)$:. Coming now 
to superoperators $\hat{\gamma}(t)$, $\hgv(t)$, $\tilde{\gamma}(t)$, 
$\tgv(t)$ with the help of inversed transformations \refp{e6.19}, one can 
obtain for the calculation of transport cores or the corresponding Green 
functions, a generalization of the Wick formulae and Feynman diagram 
technique on a nonequilibrium case. It is one of important features of the 
nonequilibrium thermo field dynamics. It consists in the possibility of 
constructing the quasiequilibrium thermo vacuum state vector as a vacuum 
state vector for the solution to the Schr\"odinger equation \refp{e2.6} of a 
quantum field system.

In the following section we consider equations of thermo field hydrodynamics 
of quantum systems of both strongly and weakly nonequilibrium states using 
the base transport equations \refp{e4.14}. We obtain expressions for 
generalized transport coefficients of viscosity and thermal conductivity in 
thermo field representation.

\section{Generalized hydrodynamic equations in thermo field representation}

In the case of description of the nonequilibrium state of a quantum system 
the following quantities can be chosen as parameters of a shortened 
description: they are density operators of particle number $n_{\bmv{k}}$, 
momentum $\bmv{p}_{\bmv{k}}$ and energy $E_{\bmv{k}}$. For such a set of 
values quasiequilibrium statistical operator $\hat{\vrhoq}(t)$ reads 
\cite{c21,c33,c34}:
%
%
{\small
\bea
\lefteqn{\ds\hat{\vrhoq}(t)=\exp\lc-\Phi(t)-
	\!\!\sum_{\bmv{k}}\ls\beta_{-\bmv{k}}(t)\hat{E}(t)-
	(\beta\bmv{v})_{\bmv{k}}(t)\hat{\bmv{p}}_{\bmv{k}}-
	\!\ls\beta\lp\mu-\frac{m}{2}V^2\rp\rs_{-\bmv{k}}(t)
	\hat{n}_{\bmv{k}}\rs\rc,}\nonumber\\
	\label{e7.1}
\eea}

\vspace*{-2ex}\noindent
where parameters $\mu_{-\bmv{k}}(t)$, $\bmv{v}_{-\bmv{k}}(t)$,
$\beta_{-\bmv{k}}(t)$ are defined from the self-consistency conditions
%
%
\be
\arraycolsep=0pt
\begin{array}{llll}
\la\la 1|\hat{n}_{\bmv{k}}&|\vrhoq(t)\ra\ra=&
	\,\la\la 1|\hat{n}_{\bmv{k}}&|\vrho(t)\ra\ra,\\
\la\la 1|\hat{\bmv{p}}_{\bmv{k}}&|\vrhoq(t)\ra\ra=&
	\,\la\la 1|\hat{\bmv{p}}_{\bmv{k}}&|\vrho(t)\ra\ra,\\
\la\la 1|\hat{E}_{\bmv{k}}&|\vrhoq(t)\ra\ra=&
	\,\la\la 1|\hat{E}_{\bmv{k}}&|\vrho(t)\ra\ra,\\
\end{array}
\label{e7.2}
\ee
and mean a chemical potential, average hydrodynamical velocity and a local
value of inversed temperature, respectively \cite{c21}. Expression
\refp{e7.1} aids to obtain a set of equations of generalized hydrodynamics 
in thermo field representation. Taking into account \refp{e4.14}, we obtain:
%
%
{
\bea
\lefteqn{\ds\ddt\la\la 1|\hat{n}_{\bmv{k}}|\vrho(t)\ra\ra=
	\la\la 1|\dot{\hat{n}}_{\bmv{k}}|\vrhoq(t)\ra\ra,}\label{e7.3}\\
\lefteqn{\ds\ddt\la\la 1|\hat{\bmv{p}}_{\bmv{k}}|\vrho(t)\ra\ra=
	\la\la 1|\dot{\hat{\bmv{p}}}_{\bmv{k}}|
	\vrhoq(t)\ra\ra-{}}\label{e7.4}\\
\lefteqn{\ds\sum_{\gb}\intl_{-\infty}^t\d t'\;\e^{\veps(t'-t)}
	\La\!\!\La J_p(\bmv{k};t)\left|T(t,t')\left|
	\mbox{$\intl_0^1$}\d\tau\;
	\vrhoq^\tau(t')J_p(\gb;t')
	\vrhoq^{1-\tau}(t')\Ra\!\!\Ra\right.\right.
	(\beta\bmv{v})_{\gb}(t')+{}}\nonumber\\
\lefteqn{\ds\sum_{\gb}\intl_{-\infty}^t\d t'\;\e^{\veps(t'-t)}
	\La\!\!\La J_p(\bmv{k};t)\left|T(t,t')\left|
	\mbox{$\intl_0^1$}\d\tau\;
	\vrhoq^\tau(t')J_E(\gb;t')
	\vrhoq^{1-\tau}(t')\Ra\!\!\Ra\right.\right.
	\beta_{\gb}(t'),}\nonumber\\
\lefteqn{\ds\ddt\la\la 1|\hat{E}_{\bmv{k}}|\vrho(t)\ra\ra=
	\la\la 1|\dot{\hat{E}}_{\bmv{k}}|
	\vrhoq(t)\ra\ra-{}}\label{e7.5}\\
\lefteqn{\ds\sum_{\gb}\intl_{-\infty}^t\d t'\;\e^{\veps(t'-t)}
	\La\!\!\La J_E(\bmv{k};t)\left|T(t,t')\left|
	\mbox{$\intl_0^1$}\d\tau\;
	\vrhoq^\tau(t')J_p(\gb;t')
	\vrhoq^{1-\tau}(t')\Ra\!\!\Ra\right.\right.
	(\beta\bmv{v})_{\gb}(t')+{}}\nonumber\\
\lefteqn{\ds\sum_{\gb}\intl_{-\infty}^t\d t'\;\e^{\veps(t'-t)}
	\La\!\!\La J_E(\bmv{k};t)\left|T(t,t')\left|
	\mbox{$\intl_0^1$}\d\tau\;
	\vrhoq^\tau(t')J_E(\gb;t')
	\vrhoq^{1-\tau}(t')\Ra\!\!\Ra\right.\right.
	\beta_{\gb}(t'),}\nonumber
\eea}
where
\[
\ba{lcllcl}
\hat{n}_{\bmv{k}}&=&n_{\bmv{k}}\lp\hav_{\bmv{k}},\hai_{\bmv{k}}\rp,\\[0.75ex]
\hat{\bmv{p}}_{\bmv{k}}&=&
	\bmv{p}_{\bmv{k}}\lp\hav_{\bmv{k}},\hai_{\bmv{k}}\rp,&\qquad
	J_p(\bmv{k};t)&=&\Big(1-\SP(t)\Big)\dot{\bmv{p}}_{\bmv{k}},\\[0.75ex]
\hat{E}_{\bmv{k}}&=&E_{\bmv{k}}\lp\hav_{\bmv{k}},\hai_{\bmv{k}}\rp,&\qquad
	J_E(\bmv{k};t)&=&\Big(1-\SP(t)\Big)\dot{E}_{\bmv{k}},\\
\ea
\]
are generalized flows. But the Mori projection operator $\SP(t)$ at the
description of the hydrodynamic state reads:
%
%
\bea
\lefteqn{\ds\SP(t)(\ldots)=\la\la 1|\ldots\vrhoq(t)\ra\ra+
	\!\!\!\!\!\!\sum_{a=\{n,\bmv{p},E\}}
	\sum_{\bmv{k}}\ls\frac{\delta\la\la 1|\ldots\vrhoq(t)}
	{\delta\la\la 1|\hat{a}_{\bmv{k}}|\vrho(t)\ra\ra}\big(a_{\bmv{k}}-
	\la\la 1|\hat{a}_{\bmv{k}}|\vrho(t)\ra\ra\big)\rs.}
	\nonumber\\
	\label{e7.6}
\eea
Transport cores
%
%
\bea
\lefteqn{\ds\La\!\!\La J_p(\bmv{k};t)
	\left|T(t,t')\left|\mbox{$\intl_0^1$}\d\tau\;
	\vrhoq^\tau(t')J_p(\gb;t')
	\vrhoq^{1-\tau}(t')\right.\right.
	\Ra\!\!\Ra,}\nonumber\\
\lefteqn{\ds\La\!\!\La J_p(\bmv{k};t)
	\left|T(t,t')\left|\mbox{$\intl_0^1$}\d\tau\;
	\vrhoq^\tau(t')J_E(\gb;t')
	\vrhoq^{1-\tau}(t')\right.\right.
	\Ra\!\!\Ra,}\nonumber\\
\lefteqn{\ds\La\!\!\La J_{\! E}(\bmv{k};t)
	\left|T(t,t')\left|\mbox{$\intl_0^1$}\d\tau\;
	\vrhoq^\tau(t')J_p(\gb;t')
	\vrhoq^{1-\tau}(t')\right.\right.
	\Ra\!\!\Ra,}\label{e7.7}\\
\lefteqn{\ds\La\!\!\La J_{\! E}(\bmv{k};t)
	\left|T(t,t')\left|\mbox{$\intl_0^1$}\d\tau\;
	\vrhoq^\tau(t')J_E(\gb;t')
	\vrhoq^{1-\tau}(t')\right.\right.
	\Ra\!\!\Ra,}\nonumber
\eea
in \refp{e7.4}, \refp{e7.5} are calculated with the help of quasiequilibrium
thermo vacuum state vector \refp{e7.1}.

For the description of nonequilibrium hydrodynamic state of a quantum field
system near equilibrium, the set of equations of generalized hydrodynamics
\refp{e7.3}--\refp{e7.5} according to \refp{e5.1}--\refp{e5.5} becomes 
closed:
%
%
\bea
\lefteqn{\ds\ddt\la\la 1|\delta\hat{n}_{\bmv{k}}|\vrho(t)\ra\ra=
	\sum_{\gb}\Omega_{n\bmv{p}}(\bmv{k},\gb)
	\la\la 1|\delta\hat{\bmv{p}}_{\gb}|\vrho(t)\ra\ra,}
	\label{e7.8}\\
\lefteqn{\ds\ddt\la\la 1|\delta\hat{\bmv{p}}_{\bmv{k}}|\vrho(t)\ra\ra=
	\sum_{\gb}\Omega_{\bmv{p}n}(\bmv{k},\gb)
	\la\la 1|\delta\hat{n}_{\gb}|\vrho(t)\ra\ra+{}}\label{e7.9}\\
&&\sum_{\gb}\Omega_{\bmv{p}h}(\bmv{k},\gb)
	\la\la 1|\delta\hat{h}_{\gb}|\vrho(t)\ra\ra-{}\nonumber\\
&&\sum_{\gb}\intl_{-\infty}^t\d t'\;\e^{\veps(t'-t)}
	\bmv{k}:\eta(\bmv{k},\gb;t,t'):\gb
	\la\la 1|\delta\hat{\bmv{p}}_{\gb}|\vrho(t')\ra\ra-{}\nonumber\\
&&\sum_{\gb}\intl_{-\infty}^t\d t'\;\e^{\veps(t'-t)}
	\bmv{k}:\xi_{\bmv{p}h}(\bmv{k},\gb;t,t')\cdot\gb
	\la\la 1|\delta\hat{h}_{\gb}|\vrho(t')\ra\ra,\nonumber\\
\lefteqn{\ds\ddt\la\la 1|\delta\hat{h}_{\bmv{k}}|\vrho(t)\ra\ra=
	\sum_{\gb}\Omega_{h\bmv{p}}(\bmv{k},\gb)
	\la\la 1|\delta\hat{\bmv{p}}_{\gb}|\vrho(t)\ra\ra-{}}
	\label{e7.10}\\
&&\sum_{\gb}\intl_{-\infty}^t\d t'\;\e^{\veps(t'-t)}
	\bmv{k}:\xi_{h\bmv{p}}(\bmv{k},\gb;t,t'):\gb
	\la\la 1|\delta\hat{\bmv{p}}_{\gb}|\vrho(t')\ra\ra-{}
	\nonumber\\
&&\sum_{\gb}\intl_{-\infty}^t\d t'\;\e^{\veps(t'-t)}
	\bmv{k}\cdot\lambda(\bmv{k},\gb;t,t')\cdot\gb
	\la\la 1|\delta\hat{h}_{\gb}|\vrho(t')\ra\ra.\nonumber
\eea
In these relations all quantities $\Omega_{n\bmv{p}}(\bmv{k},\gb)$,
$\Omega_{\bmv{p}n}(\bmv{k},\gb)$, $\Omega_{\bmv{p}h}(\bmv{k},\gb)$
and $\Omega_{h\bmv{p}}(\bmv{k},\gb)$ are defined by the relation
\refp{e5.6} and are equilibrium quantum correlation functions;
$\eta(\bmv{k},\gb;t,t')$, $\lambda(\bmv{k},\gb;t,t')$,
$\xi_{\bmv{p}h}(\bmv{k},\gb;t,t')$,
$\xi_{h\bmv{p}}(\bmv{k},\gb;t,t')$ are, correspondingly,
time-de\-pen\-dent generalized transport coefficients of viscosity, thermal
conductivity and cross transport coefficients. They are defined as follows:
%
%
\bea
\bmv{k}:\eta(\bmv{k},\gb;t,t'):\gb&=&
\La\!\!\La J_{\bmv{p}}(\bmv{k})T_0(t,t')\left|\mbox{$\intl_0^1$}\d\tau\;
	\vrho_0^\tau J_{\bmv{p}}(\gb)\vrho_0^{1-\tau}\right.\Ra\!\!\Ra
	\Chi_{\bmv{p}\bmv{p}}^{-1}(\gb),\label{e7.11}\\
\bmv{k}:\xi_{\bmv{p}h}(\bmv{k},\gb;t,t')\cdot\bmv{k}&=&
\La\!\!\La J_{\bmv{p}}(\bmv{k})T_0(t,t')\left|\mbox{$\intl_0^1$}\d\tau\;
	\vrho_0^\tau J_{h}(\gb)\vrho_0^{1-\tau}\right.\Ra\!\!\Ra
	\Chi_{hh}^{-1}(\gb),\label{e7.12}\\
\bmv{k}\cdot\xi_{h\bmv{p}}(\bmv{k},\gb;t,t'):\gb&=&
\La\!\!\La J_{h}(\bmv{k})T_0(t,t')\left|\mbox{$\intl_0^1$}\d\tau\;
	\vrho_0^\tau J_{\bmv{p}}(\gb)\vrho_0^{1-\tau}\right.\Ra\!\!\Ra
	\Chi_{\bmv{p}\bmv{p}}^{-1}(\gb),\label{e7.13}\\
\bmv{k}\cdot\lambda(\bmv{k},\gb;t,t')\cdot\gb&=&
\La\!\!\La J_{h}(\bmv{k})T_0(t,t')\left|\mbox{$\intl_0^1$}\d\tau\;
	\vrho_0^\tau J_{h}(\gb)\vrho_0^{1-\tau}\right.\Ra\!\!\Ra
	\Chi_{hh}^{-1}(\gb),\label{e7.14}
\eea
where
%
%
\bea
J_{\bmv{p}}(\bmv{k})&=&(1-\SP_0)\dot{\bmv{p}}_{\bmv{k}},\nonumber\\
J_{h}(\bmv{k})&=&(1-\SP_0)\dot{h}_{\bmv{k}},\qquad
	h_{\bmv{k}}=E_{\bmv{k}}-\Chi_{En}(\bmv{k})\Chi^{-1}_{nn}(\bmv{k})
	n_{\bmv{k}},\label{e7.15}
\eea
and $\Chi_{\bmv{p}\bmv{p}}(\bmv{k})$, $\Chi_{hh}(\bmv{k})$,
$\Chi_{En}(\bmv{k})$, $\Chi_{nn}(\bmv{k})$ are equilibrium quantum
correlation functions:
\[
\ba{lcllcl}
\Chi_{\bmv{p}\bmv{p}}(\gb)&=&\ds\La\!\!\La\bmv{p}_{\gb}\left|
	\mbox{$\intl_0^1$}\d\tau\;
	\vrho^\tau_0\bmv{p}_{-\gb}\vrho^{1-\tau}_0
	\right.\Ra\!\!\Ra,&\quad
	\Chi_{hh}(\gb)&=&\ds\La\!\!\La h_{\gb}\left|
	\mbox{$\intl_0^1$}\d\tau\;\vrho^\tau_0 h_{-\gb}\vrho^{1-\tau}_0
	\right.\Ra\!\!\Ra,\\
\Chi_{nn}(\gb)&=&\ds\La\!\!\La n_{\gb}\left|
	\mbox{$\intl_0^1$}\d\tau\;\vrho^\tau_0 n_{-\gb}\vrho^{1-\tau}_0
	\right.\Ra\!\!\Ra,&\quad
	\Chi_{En}(\gb)&=&\ds\La\!\!\La\! E_{\gb}\left|
	\mbox{$\intl_0^1$}\d\tau\;\vrho^\tau_0 n_{-\gb}\vrho^{1-\tau}_0
	\right.\Ra\!\!\Ra.\\
\ea
\]
In \refp{e7.15} the Mori projection operator $\SP_0$ is defined in 
accordance with \refp{e5.8} and is built on operators $n_{\bmv{k}}$, 
$\bmv{p}_{\bmv{k}}$, $h_{\bmv{k}}$ in thermo field representation. Operator
$h_{\bmv{k}}$ appears due to the inclusion into $|\vrhoq(t)\ra\ra$ 
\refp{e7.1} parameters $\beta_{\bmv{k}}(t)$ and $\mu_{\bmv{k}}(t)$ with the 
help of self-consistency conditions \refp{e7.2} in the linear approximation 
(see \refp{e5.1}--\refp{e5.4}) and has the meaning of generalized enthalpy. 
On the basis of hydrodynamic equations \refp{e7.8}--\refp{e7.10} one can 
investigate quantum correlation functions like ``density-density'' 
$\Phi_{nn}(\bmv{k};t)$, ``flow-flow'' $\Phi_{\bmv{p}\bmv{p}}(\bmv{k};t)$,
``enthalpy-enthalpy'' $\Phi_{hh}(\bmv{k};t)$ and the corresponding Green
functions for specific quantum field systems.

A separate and very important problem is the calculation of transport
coefficients \refp{e7.11}--\refp{e7.14}. Consideration of nonequilibrium
thermodynamics in a field theory on the basis of nonequilibrium statistical
operator \cite{c33} and approximate calculation of viscosity and thermal
conductivity coefficients for the $\mPhi^4$ field model were done in paper 
\cite{c43}. These calculations were carried out using the Green-Kubo 
formulae which connect transport coefficients of viscosity and thermal 
conductivity with the corresponding Green functions. The last ones are
built on the stress tensor and energy flow operators. In their turn, the 
Green functions were calculated using the Dzyaloshinski diagram method 
\cite{c44}.

\section{Transport equations of a consistent description of the 
\protect\newline kinetics and hydrodynamics of dense quantum systems}

In the studies of nonequilibrium states of quantum field systems, such as a
nuclear matter \cite{c8,c9}, there arises a problem of taking into
consideration coupled states. Kinetic and hydrodynamic processes in
a hot, compressed nuclear matter, which appears after ultrarelativistic
collisions of heavy nuclei or laser thermonuclear synthesis, are mutually
connected and we should consider coupled states between nuclons. This is of
great importance for the analysis and correlation of final reaction
products. Obviously, a nuclon interaction investigation based on a
quark-gluon plasma is a sequential microscopic approach to the dynamical
description of reactions in a nuclear matter. The problem of a consistent
description of  the kinetics and hydrodynamics of a dense quark-gluon 
plasma is considered in section~9. For the description of kinetic processes 
in a nuclear matter on the level of model interactions, the 
Vlasov-Ueling-Uhlenbeck kinetic equation is used. This equation is used 
mainly in the case of low densities. The problems of a dense quark-gluon
matter were discussed in detail in \cite{c8,c9,c20,c45,c46,c47}.

We will consider a quantum field system in which coupled states can
appear between the particles. Let us introduce annihilation and creation
operators of a coupled state $(A\alpha)$ with $A$-particle:
%
%
\be
\ba{l}
\ds\ai_{A\alpha}(\bmv p)=\sum_{1,\ldots,A}\mPsi_{A\alpha\bmv p}
	(1,\ldots,A)a(1)\ldots a(A),\\
\ds\av_{A\alpha}(\bmv p)=\sum_{1,\ldots,A}\mPsi^*_{A\alpha\bmv p}
	(1,\ldots,A)\av(1)\ldots\av(A),\\
\ea\label{e8.1}
\ee
where $\mPsi_{A\alpha\bmv p}(1,\ldots,A)$ is a self-function of the 
$A$-particle coupled state, $\alpha$ denotes internal quantum numbers 
(spin, etc.), $\bmv{p}$ is a particle momentum, the sum covers the 
particles. Annihilation and creation operators $a(j)$ and $\av(j)$ satisfy 
the following commutation relations:
%
%
\be
[a(l),a^+(j)]_{\sigma}=\delta_{l,j},\qquad
	[a(l),a(j)]_{\sigma}=[a^+(l),a^+(j)]_{\sigma}=0,
	\label{e8.2}
\ee
where $\sigma$-commutator is determined by $[a,b]_\sigma=ab-\sigma ba$ with
$\sigma=\pm 1$: $+1$ for bosons and $-1$ for fermions.

The Hamiltonian of such a system can be written in the form:
%
%
\bea
\lefteqn{\ds H=\sum_{A,\alpha}\int\frac{\d\bmv{p}\d\bmv{q}}
	{(2\pi\hbar)^6}\frac{p^2}{2m_A}
	\av_{A\alpha}\lp\bmv{p}-\frac{\bmv{q}}{2}\rp
	\ai_{A\alpha}\lp\bmv{p}+\frac{\bmv{q}}{2}\rp+{}}
	\label{e8.3}\\
\lefteqn{\ds\frac 12\sum_{A,B}\sum_{\alpha,\beta}
	\int\frac{\d\bmv{p}\d\bmv{p}'\d\bmv{q}}{(2\pi\hbar)^9}V_{AB}(\bmv{q})
	\av_{A\alpha}\lp\bmv{p}+\frac{\bmv{q}-\bmv{p}'}{2}\rp
	\hat{n}_{B\beta}^{\php}(\bmv{q})
	\ai_{A\alpha}\lp\bmv{p}-\frac{\bmv{q}-\bmv{p}'}{2}\rp,}
	\nonumber
\eea
where $V_{AB}(\bmv{q})$ is interaction energy between $A$- and
$B$-particle coupled states, $\bmv{q}$ is a wavevector.
Annihilation and creation operators
$\ai_{A\alpha}(\bmv{p})$ and $\av_{A\alpha}(\bmv{p})$ satisfy the following 
commutation relations:
%
%
\be
\ba{l}
\ds [\ai_{A\alpha}(\bmv{p}),\av_{B\beta}(\bmv{p}')]_{\sigma}=
	\delta_{A,B}\delta_{\alpha,\beta}\delta(\bmv{p}-\bmv{p}'),\\
\ds [\ai_{A\alpha}(\bmv{p}),\ai_{B\beta}(\bmv{p}')]_{\sigma}=
	[\av_{A\alpha}(\bmv{p}),\av_{B\beta}(\bmv{p}')]_{\sigma}=0.\\
\ea\label{e8.4}
\ee
$\hat{n}_{B\beta}(\bmv{q})$ in (\ref{e8.3}) is a Fourier transform of
the $B$-particle density operator:
\[
\hat{n}_{B\beta}(\bmv{q})=\int\frac{\d\bmv{p}}{(2\pi\hbar)^3}\,
	\av_{\bmv{p}-\frac{\bmv{q}}{2}}\ai_{\bmv{p}+\frac{\bmv{q}}{2}}.
\]

As parameters of a shortened description for the consistent description of
the kinetics and hydrodynamics of a system, where coupled states between 
the particles can appear, let us choose nonequilibrium distribution 
functions of $A$-particle coupled states in thermo field representation
%
%
\be
\la\la 1|\hat{n}_{A\alpha}^{\php}(\bmv{r},\bmv{p})|\vrho(t)\ra\ra=
	f_{A\alpha}^{\php}(\bmv{r},\bmv{p};t)=f_{A\alpha}^{\php}(x;t),\quad
	x=\{\bmv{r},\bmv{p}\},\label{e8.5}
\ee
here $f_{A\alpha}^{\php}(x;t)$ is a Wigner function of the $A$-particle 
coupled state where
%
%
\be
\hat{n}_{A\alpha}^{\php}(\bmv{r},\bmv{p})\equiv\hat{n}_{A\alpha}^{\php}(x)=
	\int\frac{\d\bmv{q}}{(2\pi\hbar)^3}\,
	\e^{-\frac{1}{\im\hbar}\bmv{q}\cdot\bmv{r}}
	\hat{a}^+_{A\alpha}\lp\bmv{p}-\frac{\bmv{q}}{2}\rp
	\hat{a}^{\phantom +}_{A\alpha}\lp\bmv{p}+\frac{\bmv{q}}{2}\rp
	\label{e8.6}
\ee
is the Klimontovich density operator; and the average value of the total 
energy density operator
%
%
\be
\la\la 1|\hat{H}(\bmv{r})|\vrho(t)\ra\ra=\la\la 1|H(\bmv{r})\vrho(t)\ra\ra.
	\label{e8.7}
\ee
By this $\int\d\bmv{r}\;H(\bmv{r})=H$, $\hat{H}(\bmv{r})$ is a 
superoperator of the total energy density which is constructed on 
annihilation and creation superoperators 
$\hat{a}_{A\alpha}^{\phantom +}(\bmv{p})$ and 
$\hat{a}^+_{A\alpha}(\bmv{p})$. The latter satisfy commutation relations
(\ref{e8.4}). Following \cite{c27}, one can rewrite quasiequilibrium
statistical operator $\hat{\vrho}_{\rm q}(t)$,
$|\vrho_{\rm q}(t)\ra\ra=\hat{\vrho}_{\rm q}(t)|1\ra\ra$ for the mentioned
parameters of a shortened description in the form:
%
%
{\small
\bea
\lefteqn{\ds\hat{\vrho}_{\rm q}(t)=\exp\lc-\Phi^*(t)-
	\int\d\bmv{r}\;\beta(\bmv{r};t)\lp\hat{H}(\bmv{r})-\sum_{A,\alpha}
	\int\frac{\d\bmv{p}}{(2\pi\hbar)^3}\,
	\mu_{A\alpha}^{\php}(x;t)\hat{n}_{A\alpha}^{\php}(x)\rp\rc,}
	\nonumber\\
	\label{e8.8}
\eea}

\vspace*{-2ex}\noindent
where Lagrange multipliers $\beta(\bmv{r};t)$ and
$\mu_{A\alpha}^{\php}(x;t)$ can be found from the self-con\-sis\-ten\-cy
conditions, correspondingly:
%
%
\bea
\la\la 1|\hat{H}(\bmv{r})|\vrho(t)\ra\ra&=&
	\la\la 1|\hat{H}(\bmv{r})|\vrho_{\rm q}(t)\ra\ra,
	\label{e8.9}\\
\la\la 1|\hat{n}_{A\alpha}^{\php}(x)|\vrho(t)\ra\ra&=&
	\la\la 1|\hat{n}_{A\alpha}^{\php}(x)|\vrho_{\rm q}(t)\ra\ra,
	\label{e8.10}
\eea
$\Phi^*(t)$ is the Massieu-Planck functional and it can be defined from the
normalization condition \refp{e3.6}:
%
%
{\small
\bea
\lefteqn{\ds\Phi^*(t)=\ln\La\!\!\La 1\left|
	\exp\lc-\!\!\int\d\bmv{r}\;\beta(\bmv{r};t)
	\lp\hat{H}(\bmv{r})-\sum_{A,\alpha}
	\int\frac{\d\bmv{p}}{(2\pi\hbar)^3}
	\mu_{A\alpha}^{\php}(x;t)\hat{n}_{A\alpha}^{\php}(x)
	\rp\rc\right.\Ra\!\!\Ra.}
	\nonumber\\
	\label{e8.11}
\eea}

\vspace*{-2ex}\noindent
Using now the general structure of nonequilibrium thermo field dynamics
\refp{e4.1}--\refp{e4.14}, one can obtain a set of generalized transport
equations for $A$-particle Wigner distribution functions and the average
interaction energy:
%
%
\bea
\lefteqn{\ds \frac{\partial}{\partial t}\la\la 1|\hat{n}_{A\alpha}^{\php}(x)
	|\vrho(t)\ra\ra=\la\la 1|\dot{\hat{n}}_{A\alpha}^{\php}(x)
	|\vrho_{\rm q}(t)\ra\ra+{}}\label{e8.12}\\
&&\int\d\bmv{r}'\int\limits_{-\infty}^t\d t'\;
	\e^{\veps(t'-t)}
	\vphi^{A\alpha}_{nH}(x,\bmv{r}';t,t')\beta(\bmv{r}';t')+{}
	\nonumber\\
&&\sum_{B,\beta}\int\d x'\int\limits_{-\infty}^t\d t'\;
	\e^{\veps(t'-t)}
	\vphi^{A\alpha B\beta}_{nn}(x,x';t,t')
	\beta(\bmv{r}';t')\mu_{B\beta}^{\php}(x';t'),
	\nonumber
\eea
%
%
%
%
%
\bea
\lefteqn{\ds \frac{\partial}{\partial t}\la\la 1|\hat{H}(\bmv{r})
	|\vrho(t)\ra\ra=\la\la 1|\dot{\hat{H}}(\bmv{r})
	|\vrho_{\rm q}(t)\ra\ra+{}}\label{e8.13}\\
&&\int\d\bmv{r}'\int\limits_{-\infty}^t\d t'\;
	\e^{\veps(t'-t)}
	\vphi_{HH}^{\phantom B}(\bmv{r},\bmv{r}';t,t')\beta(\bmv{r}';t')+{}
	\nonumber\\
&&\sum_{B,\beta}\int\d x'\int\limits_{-\infty}^t\d t'\;
	\e^{\veps(t'-t)}
	\vphi^{B\beta}_{Hn}(\bmv{r},x';t,t')
	\beta(\bmv{r}';t')\mu_{B\beta}^{\php}(x';t'),
	\nonumber
\eea
where $x'=\{\bmv{r}',\bmv{p}'\}$,
$dx'=(2\pi\hbar)^{-3}d\bmv{r}'\,d\bmv{p}'$. Here
%
%
\bea
\vphi_{nn}^{A\alpha\atop B\beta}(x,x';t,t')&=&
	\La\!\!\La 1\bigg|\hat{J}_{n_{A\alpha}^{\php}}(x,t)T(t,t')
	\bigg|\mbox{$\intl_0^1$}\d\tau\;\vrho_{\rm q}^\tau(t')
	J_{n_{B\beta}^{\php}}(x';t')\vrho_{\rm q}^{1-\tau}(t')\!\Ra\!\!\Ra\!,
	\label{e8.14}\\
\vphi_{nH}^{A\alpha}(x,\bmv{r}';t,t')&=&
	\La\!\!\La 1\bigg|\hat{J}_{n_{A\alpha}^{\php}}(x,t)T(t,t')
	\bigg|\mbox{$\intl_0^1$}\d\tau\;\vrho_{\rm q}^\tau(t')
	J_{H}(\bmv{r}';t')\vrho_{\rm q}^{1-\tau}(t')\Ra\!\!\Ra,
	\label{e8.15}\\
\vphi_{Hn}^{B\beta}(\bmv{r}',x';t,t')&=&
	\La\!\!\La 1\bigg|\hat{J}_{H}(\bmv{r},t)T(t,t')
	\bigg|\mbox{$\intl_0^1$}\d\tau\;\vrho_{\rm q}^\tau(t')
	J_{n_{B\beta}^{\php}}(x';t')\vrho_{\rm q}^{1-\tau}(t')\Ra\!\!\Ra,
	\label{e8.16}\\
\vphi_{HH}^{\phantom B}(\bmv{r},\bmv{r}';t,t')&=&
	\La\!\!\La 1\bigg|\hat{J}_{H}(\bmv{r},t)T(t,t')
	\bigg|\mbox{$\intl_0^1$}\d\tau\;\vrho_{\rm q}^\tau(t')
	J_{H}(\bmv{r}';t')\vrho_{\rm q}^{1-\tau}(t')\Ra\!\!\Ra,
	\label{e2.30}
\eea
are generalized transport cores which describe dissipative processes. In 
these formulae
%
%
\be
\begin{array}{@{}lll}
\ds J_H(\bmv{r};t)&=&\ds\Big(1-{\SP}(t')\Big)\dot{H}(\bmv{r}),\\[0.75ex]
\ds J_{n_{A\alpha}^{\php}}(\bmv{r},\bmv{p};t)&=&\ds\Big(1-{\SP}(t')\Big)
	\dot{n}_{A\alpha}^{\php}(x)\\
\end{array}
\label{e8.18}
\ee
are generalized flows,
\[
\begin{array}{@{}lll}
\ds\dot{H}(\bmv{r})&=&\ds-\frac{1}{\im\hbar}[H,H(\bmv{r})],\\[1.5ex]
\ds\dot{n}_{A\alpha}^{\php}(\bmv{r},\bmv{p})&=&\ds
	-\frac{1}{\im\hbar}[H,n_{A\alpha}^{\php}(x)],\\
\end{array}
\]
${\SP}(t)$ is a generalized Mori projection operator in thermo field
representation. It acts on operators
%
%
\bea
\lefteqn{\ds{\SP}(t)P=\la\la|\hat{P}|\vrho_{\rm q}(t)\ra\ra+{}}
\label{e8.19}\\
&&\int\d\bmv{r}\;\frac{\delta\la\la 1|\hat{P}|\vrho_{\rm q}(t)\ra\ra}
	{\delta\la\la 1|\hat{H}(\bmv{r})|\vrho(t)\ra\ra}
	\Big(H(\bmv{r})-\la\la 1|\hat{H}(\bmv{r})|\vrho(t)\ra\ra\Big)+{}
	\nonumber\\
&&\sum_{A,\alpha}\int\frac{\d\bmv{r}\,\d\bmv{p}}{(2\pi\hbar)^3}
	\frac{\delta\la\la 1|\hat{P}|\vrho_{\rm q}(t)\ra\ra}
	{\delta\la\la 1|\hat{n}_{A\alpha}^{\php}(x)|\vrho(t)\ra\ra}
	\Big(n_{A\alpha}^{\php}(x)-
	\la\la 1|\hat{n}_{A\alpha}^{\php}(x)|\vrho(t)\ra\ra\Big)
	\nonumber
\eea
and has all the properties of a projection operator:
\[
\begin{array}{@{}lll@{\qquad}lll}
\ds{\SP}(t)H(\bmv{r})&=&H(\bmv{r}),&
	\ds{\SP}(t){\SP}(t')&=&{\SP}(t),\\[0.75ex]
\ds{\SP}(t)n_{A\alpha}^{\php}(\bmv{r},\bmv{p})&=&
	\ds n_{A\alpha}^{\php}(\bmv{r},\bmv{p}),&
	\ds\Big(1-{\SP}(t)\Big){\SP}(t)&=&0.
\end{array}
\]

The obtained transport equations have the general meaning and can describe 
both weakly and strongly nonequilibrium processes of a quantum system with 
taking into consideration coupled states. In a low density quantum field
Bose- or Fermi-system the influence of the average value of interaction 
energy is substantially smaller than the average kinetic energy, and 
coupled states between the particles are absent. In such a case the set of 
transport equations \refp{e8.12}, \refp{e8.13} is simplified. It transforms 
into a kinetic equation \cite{c27} in thermo field representation for the 
average value of the Klimontovich operator 
$\la\la 1|\hat{n}(x)|\vrho(t)\ra\ra$:
{\small
\bean
\lefteqn{\ds\ddt\la\la 1|\hat{n}_{\bmv{k}}(\bmv{p})|\vrho(t)\ra\ra=
	\la\la 1|\dot{\hat{n}}_{\bmv{k}}(\bmv{p})|
	\vrhoq(t)\ra\ra+{}}\\
\lefteqn{\ds\sum_{\bmv{g}}\int\d\bmv{p}'
	\intl_{-\infty}^t\d t'\;\e^{\veps(t'-t)}
	\La\!\!\La J_n(\bmv{k};t)\left|T(t,t')\left|
	\mbox{$\intl_0^1$}\d\tau\;
	\vrhoq^\tau(t')J_n(\bmv{g};t')
	\vrhoq^{1-\tau}(t')\Ra\!\!\Ra\right.\right.
	b_{-\bmv{g}}(\bmv{p}';t').}\nonumber
\eean}
Using the projection operators method, this equation was obtained in 
\cite{c24}.

In the next step we will construct such annihilation and creation 
superoperators, for which the quasiequilibrium thermo vacuum state vector 
is a vacuum state. Analysing the structure of quasiequilibrium statistical
superoperator \refp{e8.8}, one can mark out some part which would
correspond to the system of noninteracting quantum $A$-particles. Let us
write $\hat{\vrho}_{\rm q}(t)$ in an evident form and separate terms which
are connected with the interaction energy between the particles:
%
%
\bea
\lefteqn{\ds\hat{\vrho}_{\rm q}(t)=\exp\lc-\Phi^*(t)-
	\int\d\bmv{r}\;\beta(\bmv{r};t)
	\times{}\right.}\label{e8.20}\\
\lefteqn{\ds\left.\sum_{A,\alpha}\int\frac{\d\bmv{r}\d\bmv{p}}{(2\pi\hbar)^3}
	\ls\frac{\bmv{p}^2}{2m_A}\hat{n}_{A\alpha}^{\php}(x)-
	\mu_{A\alpha}^{\php}(x;t)\hat{n}_{A\alpha}^{\php}(x)\rs-
	\int\d\bmv{r}\beta(\bmv{r};t)\hat{H}_{\rm int}(\bmv{r})\right\}.}
	\nonumber
\eea
Using operator equality ($A$ and $B$ are some operators)
\[
\ds\e^{A+B}=\ls 1+\int\limits_0^1\d\tau\;
	\e^{\tau(A+B)}\,B\,\e^{-\tau(A+B)}\rs\e^{A},
\]
the relation for $\hat{\vrho}_{\rm q}(t)$ can be rewritten in the following
form:
%
%
\be
\hat{\vrho}_{\rm q}(t)=\ls 1-\int\d\bmv{r}\;\beta(\bmv{r};t)
	\int\limits_0^1\d\tau\;
	\hat{\vrho}_{\rm q}^\tau(t)\hat{H}_{\rm int}(\bmv{r})
	\big(\hat{\vrho}_{\rm q}(t)\big)^{-\tau}\rs
	\hat{\vrho}_{\rm q}^0(t),\label{e8.21}
\ee
where
%
%
\bea
\lefteqn{\ds\hat{\vrho}_{\rm q}^0(t)=\exp\lc\Phi(t)-
	\int\d\bmv{r}\;\beta(\bmv{r};t)
	\times{}\right.}\label{e8.22}\\
&&\left.\sum_{A,\alpha}\int\frac{\d\bmv{r}\,\d\bmv{p}}{(2\pi\hbar)^3}\ls
	\frac{p^2}{2m_A}\hat{n}_{A\alpha}^{\php}(x)-
	\mu_{A\alpha}^{\php}(x;t)\hat{n}_{A\alpha}^{\php}(x)\rs\rc,
	\nonumber
\eea
or
%
%
\be
\hat{\vrho}_{\rm q}^0(t)=\exp\lc\Phi(t)-
	\int\d\bmv{r}\;\beta(\bmv{r};t)
	\sum_{A,\alpha}\int\frac{\d\bmv{p}}{(2\pi\hbar)^3}
	b_{A\alpha}^{\php}(x;t)\hat{n}_{A\alpha}^{\php}(x)\rc,
	\label{e8.23}
\ee
where $\ds b_{A\alpha}^{\php}(x;t)=
\ls\frac{\bmv{p}^2}{2m_A}\hat{n}_{A\alpha}^{\php}(x)-
\mu_{A\alpha}^{\php}(x;t)\hat{n}_{A\alpha}^{\php}(x)\rs.$
Quasiequilibrium statistical superoperator $\hat{\vrho}_{\rm q}^0(t)$
is bilinear on annihilation and creation superoperators
$\hai_{A\alpha}(\bmv{P})$ and $\hav_{A\alpha}(\bmv{P})$, as well as on the
non-perturbed part of Hamiltonian $\bar{H}_0$. One can write the total
quasiequilibrium superoperator as some non-perturbed part of
$\hat{\vrho}_{\rm q}^0(t)$ and the part which describes interaction of
quantum particles in the quasiequilibrium state. Further, we introduce the
following designation:
%
%
\be
\hat{\vrho}_{\rm q}(t)=\hat{\vrho}_{\rm q}^0(t)+\hat{\vrho}_{\rm q}^\prime(t),
\label{e8.24}
\ee
where
%
%
\be
\hat{\vrho}_{\rm q}^\prime(t)=-\int\d\bmv{r}\;\beta(\bmv{r};t)
	\int\limits_0^1\d\tau\;
	\hat{\vrho}_{\rm q}^\tau(t)\hat{H}_{\rm int}(\bmv{r})
	\big(\hat{\vrho}_{\rm q}(t)\big)^{-\tau}\hat{\vrho}_{\rm q}^0(t).
	\label{e8.25}
\ee
Quasiequilibrium thermo vacuum states $|\hat{\vrho}_{\rm q}(t)\ra\ra$ and
$|\hat{\vrho}_{\rm q}^0(t)\ra\ra$ are not vacuum states for annihilation and
creation superoperators $\hai_{A\alpha}(\bmv{P})$,
$\hav_{A\alpha}(\bmv{P})$, $\tai_{A\alpha}(\bmv{P})$,
$\tav_{A\alpha}(\bmv{P})$. But for $|\hat{\vrho}_{\rm q}^0(t)\ra\ra$ one can
construct new superoperators $\hgi_{A\alpha}(\bmv{P})$,
$\hgv_{A\alpha}(\bmv{P})$, $\tgi_{A\alpha}(\bmv{P})$,
$\tgv_{A\alpha}(\bmv{P})$ as a linear combination of  superoperators
$\hai_{A\alpha}(\bmv{P})$, $\hav_{A\alpha}(\bmv{P})$ and
$\tai_{A\alpha}(\bmv{P})$, $\tav_{A\alpha}(\bmv{P})$ in order to satisfy 
the conditions:
%
%
\be
\ba{lll}
\ds\hgi_{A\alpha}(\bmv{P};t)|\vrho_{\rm q}^0(t)\ra\ra=0,
	&\qquad&\ds\la\la 1|\hgv_{A\alpha}(\bmv{P};t)=0,\\[0.5ex]
\ds\tgi_{A\alpha}(\bmv{P};t)|\vrho_{\rm q}^0(t)\ra\ra=0,
	&\qquad&\ds\la\la 1|\tgv_{A\alpha}(\bmv{P};t)=0.\\
\ea\label{e8.26}
\ee
To achieve this let us consider an action of annihilation superoperators
$\hai_{A\alpha}(\bmv{P};t)$, $\tai_{A\alpha}(\bmv{P};t)$ on quasiequilibrium
state $|\vrho_{\rm q}^0(t_0)\ra\ra$:
%
%
\bea
\ds\hai_{A\alpha}(\bmv{P};t)|\vrho_{\rm q}^0(t_0)\ra\ra&=&
	\phantom{\sigma}f_{A\alpha}(\bmv{P};t-t_0)
	\tav_{A\alpha}(\bmv{P};t)|\vrho_{\rm q}^0(t_0)\ra\ra,
	\nonumber\\
\ds\tai_{A\alpha}(\bmv{P};t)|\vrho_{\rm q}^0(t_0)\ra\ra&=&
	\sigma f_{A\alpha}(\bmv{P};t-t_0)
	\hav_{A\alpha}(\bmv{P};t)|\vrho_{\rm q}^0(t_0)\ra\ra,
	\label{e8.27}
\eea
where superoperators $\hai_{A\alpha}(\bmv{p};t)$,
$\hav_{A\alpha}(\bmv{p};t)$,
$\tai_{A\alpha}(\bmv{p};t)$, $\hav_{A\alpha}(\bmv{p};t)$ are in the
Heisenberg representation
\bean
\lefteqn{\ds\hai_{A\alpha}(\bmv{P};t)=
	\e^{-\frac{1}{\im\hbar}\bar{H}_0t}
	\,\hai_{A\alpha}(\bmv{P})\,
	\e^{\frac{1}{\im\hbar}\bar{H}_0t},\qquad
	\tai_{A\alpha}(\bmv{P};t)=
	\e^{-\frac{1}{\im\hbar}\bar{H}_0t}
	\,\tai_{A\alpha}(\bmv{P})\,
	\e^{\frac{1}{\im\hbar}\bar{H}_0t},}\\
\lefteqn{\ds\hav_{A\alpha}(\bmv{P};t)=
	\e^{-\frac{1}{\im\hbar}\bar{H}_0t}
	\,\hav_{A\alpha}(\bmv{P})\,
	\e^{\frac{1}{\im\hbar}\bar{H}_0t},\qquad
	\tav_{A\alpha}(\bmv{P};t)=
	\e^{-\frac{1}{\im\hbar}\bar{H}_0t}
	\,\tav_{A\alpha}(\bmv{P})\,
	\e^{\frac{1}{\im\hbar}\bar{H}_0t},}
\eean
and satisfy commutation relations:
\[
\ba{l}
\ds\ls\hai_{A\alpha}(\bmv{P};t),\hav_{B\beta}(\bmv{P}';t)\rs_\sigma=
	\delta_{A,B}\delta_{\alpha,\beta}\delta(\bmv{P}-\bmv{P}'),\\
\ds\ls\tai_{A\alpha}(\bmv{P};t),\tav_{B\beta}(\bmv{P}';t)\rs_\sigma=
	\delta_{A,B}\delta_{\alpha,\beta}\delta(\bmv{P}-\bmv{P}'),\\
\ds\ls\hai_{A\alpha}(\bmv{P};t),\tai_{B\beta}(\bmv{P}';t)\rs_\sigma=
	\ls\hav_{A\alpha}(\bmv{P};t),\tav_{B\beta}(\bmv{P}';t)\rs_\sigma=0.\\
\ea
\]
It is necessary to note that superoperators $\hat{H}(\bmv{r})$,
$\hat{n}_{A\alpha}^{\php}(x)$ are built on superoperators
$\hai_{A\alpha}(\bmv{p}+\frac{\bmv{q}}{2})$,
$\hav_{A\alpha}(\bmv{p}-\frac{\bmv{q}}{2})$,
$\tai_{A\alpha}(\bmv{p}+\frac{\bmv{q}}{2})$,
$\tav_{A\alpha}(\bmv{p}-\frac{\bmv{q}}{2})$.
Therefore, for convenience here a unit denotion was introduced for arguments
like $\bmv{P}=\bmv{p}\pm\frac{\bmv{q}}{2}$. This should be taken into
account in further calculations where obvious expressions are needed.

According to general relations of section 6 \refp{e6.7}--\refp{e6.19},
we can introduce new operators
$\hgi_{A\alpha}(\bmv{P};t)$, $\hgv_{A\alpha}(\bmv{P};t)$,
$\tgi_{A\alpha}(\bmv{P};t)$, $\tgv_{A\alpha}(\bmv{P};t)$ via
superoperators
$\hai_{A\alpha}(\bmv{P};t)$, $\hav_{A\alpha}(\bmv{P};t)$,
$\tai_{A\alpha}(\bmv{P};t)$, $\tav_{A\alpha}(\bmv{P};t)$:
%
%
\bea
\ds\hgi_{A\alpha}(\bmv{P};t)&=&
	\sqrt{1+\sigma n_{A\alpha}^{\php}(\bmv{P};t,t_0)}
	\ls\hai_{A\alpha}(\bmv{P};t)-
	{\frac{n_{A\alpha}^{\php}(\bmv{P};t,t_0)}
	{1+\sigma n_{A\alpha}^{\php}(\bmv{P};t,t_0)}}
	\tav_{A\alpha}(\bmv{P};t)\rs,\nonumber\\
\ds\tgv_{A\alpha}(\bmv{P};t)&=&
	\sqrt{1+\sigma n_{A\alpha}^{\php}(\bmv{P};t,t_0)}
	\ls\tav_{A\alpha}(\bmv{P};t)-
	\sigma\hai_{A\alpha}(\bmv{P};t)\rs.
	\label{e8.28}
\eea
Relations \refp{e8.28} satisfy conditions (8.26). Here
\bean
n_{A\alpha}^{\php}(\bmv{p},\bmv{q};t,t_0)=
	n_{A\alpha}^{\php}(\bmv{P};t,t_0),=\nonumber
\lefteqn{\ds\la\la 1|\tav_{A\alpha}(\bmv{P};t)\tai_{A\alpha}(\bmv{P};t)
	|\vrho_{\rm q}^0(t_0)\ra\ra={}}\\
\lefteqn{\ds\la\la 1|\tav_{A\alpha}(\bmv{p}-\frac{\bmv{q}}{2};t)
	\tai_{A\alpha}(\bmv{p}+\frac{\bmv{q}}{2};t)|\vrho_{\rm q}^0(t_0)
	\ra\ra,}
\eean
is a quasiequilibrium distribution function of $A$-particle coupled
states in momentum space $\bmv{p}$, $\bmv{q}$, which is calculated with the
help of quasiequilibrium thermo vacuum state vector
$\vrho_{\rm q}^0(t_0)\ra\ra$ \refp{e8.23}. Function
$f_{A\alpha}(\bmv{P};t-t_0)$ in formulae \refp{e8.27} is connected
with $n_{A\alpha}(\bmv{P};t,t_0)$ by the relation
\[
f_{A\alpha}(\bmv{P};t-t_0)=\frac{n_{A\alpha}(\bmv{P};t,t_0)}
	{1+\sigma n_{A\alpha}(\bmv{P};t,t_0)}.
\]
Superoperators
$\hgi_{A\alpha}(\bmv{P};t)$ and $\tgi_{A\alpha}(\bmv{P};t)$,
$\hgv_{A\alpha}(\bmv{P};t)$ and $\tgv_{A\alpha}(\bmv{P};t)$ satisfy the
``ca\-no\-ni\-cal'' commutation relations:
%
%
\be
\ba{l}
\ds\ls\hgi_{A\alpha}(\bmv{P};t),\hgv_{B\beta}(\bmv{P}';t)\rs_\sigma=
	\delta_{A,B}\delta_{\alpha,\beta}\delta(\bmv{P}-\bmv{P}'),\\
\ds\ls\tgi_{A\alpha}(\bmv{P};t),\tgv_{B\beta}(\bmv{P}';t)\rs_\sigma=
	\delta_{A,B}\delta_{\alpha,\beta}\delta(\bmv{P}-\bmv{P}'),\\
\ds\ls\hgi_{A\alpha}(\bmv{P};t),\tgi_{B\beta}(\bmv{P}';t)\rs_\sigma=
	\ls\hgv_{A\alpha}(\bmv{P};t),\tgv_{B\beta}(\bmv{P}';t)\rs_\sigma=0.\\
\ea\label{e8.29}
\ee
Inversed transformations to superoperators $\hai_{A\alpha}(\bmv{P};t)$,
$\tav_{A\alpha}(\bmv{P};t)$ are easily obtained from \refp{e8.28}:
%
%
\bea
\ds\hai_{A\alpha}(\bmv{P};t)&=&
	\sqrt{1+\sigma n_{A\alpha}^{\php}(\bmv{P};t,t_0)}
	\ls\hgi_{A\alpha}(\bmv{P};t)+
	{\frac{n_{A\alpha}^{\php}(\bmv{P};t,t_0)}
	{1+\sigma n_{A\alpha}^{\php}(\bmv{P};t,t_0)}}
	\tgv_{A\alpha}(\bmv{P};t)\rs,\nonumber\\
\ds\tav_{A\alpha}(\bmv{P};t)&=&
	\sqrt{1+\sigma n_{A\alpha}^{\php}(\bmv{P};t,t_0)}
	\ls\tgv_{A\alpha}(\bmv{P};t)+
	\sigma\hgi_{A\alpha}(\bmv{P};t)\rs.
	\label{e8.30}
\eea
$\hgi_{A\alpha}(\bmv{P};t)$, $\hgv_{A\alpha}(\bmv{P};t)$,
$\tgi_{A\alpha}(\bmv{P};t)$, $\tgv_{A\alpha}(\bmv{P};t)$ could be
de\-fi\-ned as some operators of annihilation and creation of
$A$-quasiparticle coupled states, for which quasiequilibrium thermo vacuum
state $|\vrho_{\rm q}^0(t_0)\ra\ra$ \refp{e8.23} is a vacuum state. In such
a way, we obtained relations of dynamical reflection of
superoperators $\hai_{A\alpha}(\bmv{P};t)$, $\hav_{A\alpha}(\bmv{P};t)$,
$\tai_{A\alpha}(\bmv{P};t)$, $\tav_{A\alpha}(\bmv{P};t)$ to new
superoperators of ``quasiparticles'' $\hgi_{A\alpha}(\bmv{P};t)$,
$\hgv_{A\alpha}(\bmv{P};t)$, $\tgi_{A\alpha}(\bmv{P};t)$,
$\tgv_{A\alpha}(\bmv{P};t)$.

A set of transport equations \refp{e8.12}, \refp{e8.13} together with
dynamical reflections \refp{e8.28}, \refp{e8.30} of superoperators in the 
thermo field space constitute the basis for a consistent description of the 
kinetics and hydrodynamics of a dense quantum system with strongly coupled
states. Both strongly and weakly nonequilibrium processes of a nuclear 
matter can be investigated using this approach, in which the particle
interaction is characterized by strongly coupled states, taking into 
account theirs nuclear nature \cite{c1,c2,c8,c9}.

It is much sequential to describe investigations of kinetic and
hydrodynamic processes of a nuclear matter on the basis of quark-gluon
interaction. The quantum relativistic theory of kinetic and hydrodynamic
processes has its own problems and experiences its impetuous formation
\cite{c1,c2,c3,c4,c5,c6}. In the next section we consider one of the 
possible ways of describing the kinetics and hydrodynamics of a quark-gluon 
plasma.

\section{Thermo field transport equations for a quark-gluon plasma}

Investigation of the nonequilibrium properties of a quark-gluon plasma --
QGP which can be created after ultrarelativistic collisions of heavy
nuclei \cite{c1,c2,c3,c4,c5,c6,c7,c8,c9,c10} or laser thermonuclear
synthesis is topical from the point of view based on the statistical
approach \cite{c8,c9,c14,c15,c48,c49,c50,c51,c52,c53,c54,c55}.  Thus, it is
important to construct kinetic and hydrodynamic equations for QGP because
such a state of a nuclear matter is characterized by high temperature, 
large density and strong interactions between quarks and gluons which are
described by chromodynamics \cite{c51,c52,c53}. At present there are certain
achievements in this direction. The classical theory of transport
processes in QGP, based on the relativistic Vlasov-Boltzmann equation, was
proposed by Heinz \cite{c54,c55}. On the basis of these works, transport
coefficients for weakly nonequilibrium QGP were studied
in \cite{c56,c57,c58}. The Lenard-Balescu-type collision integral for the 
classical model of a quark plasma was obtained in \cite{c59} using the
Klimontovich method \cite{c60,c61}. Hydrodynamics of QGP is 
considered in papers \cite{c49,c50,c62,c63}. The Vlasov-Boltzmann
equations, like the quantum kinetic ones for Wigner distribution functions
for quarks and gluons, were obtained in \cite{c49,c64,c65,c66}. Some
interesting results were obtained in paper \cite{c67} where the
temperature behaviour of the kinetic coefficients of a gluon gas had been
studied using the Green-Kubo formalism and $\mPhi^4$--model.

However, it is necessary to note that the Vlasov-Boltzmann kinetic
equation is correct for a rarefied plasma only (small densities). Thus, it 
is only the first step in the investigation of transport processes of QGP.

In a dense high temperature quark-gluon plasma, which is characterized by
strong interactions, kinetic and hydrodynamic processes are mutually
connected and should be considered consistently. In this section we carry
out a consistent description of the kinetics and hydrodynamics for QGP
on the basis of a nonequilibrium thermo field dynamics using the method of
nonequilibrium statistical operator \cite{c27,c33,c34}. We will obtain
generalized relativistic quantum transport equations of the consistent
description of the kinetics and hydrodynamics for QGP. It should be also 
noted that problems of the description of nonequilibrium properties of QGP 
were considered in papers \cite{c29,c68,c69}.

Consider QGP with the Lagrangian from quantum hydrodynamics
\cite{c51,c52,c53},
%
%
\bea
L&=&\frac 14 F^a_{\mu\nu } F^a_{\mu\nu }+{\mit \Psi}^+
\lp\im\partial^\mu+\frac g2\lambda^a A^a_\mu
\gamma_\mu \rp {\mit \Psi},\label{e9.1}\\
F^a_{\mu\nu}&=&\partial_\mu A^a_\nu - \partial_\nu A^a_\mu+
g f^{abc} A^b_\mu A^c_\nu,\nonumber
\eea
where the fields of the matter are spinor quark fields ${\mit\Psi}$ for
which one employs the collective designation ${\mit \Psi} (x)$ with
components ${\mit \Psi}_\alpha^{f_i}$, where $i=1,2,3$ (index of colour:
red, green, yellow); $f=1,\ldots,6$ (flavour indices: $b$, $c$, $d$, $s$,
$t$, $u$); $\alpha=1,\ldots,4$ (spinor indices); $A^a_\nu$ are
gauge vector fields (Jang-Mill's fields) that correspond to gluons;
$a=1,\ldots,8$ indices of colour; $\lambda^a$ means  eight  Gell-Mann
matrices satisfying the commutation relations
\bean
\ls \frac{\lambda^a}{2},\frac{\lambda^b}{2}\rs =
\im f^{abc}\frac{\lambda^c}{2},
\eean
$f^{abc}$ are the structural constants of groups SU(3); $g$ is the gauge
constant of connection; $\partial_\mu=\partial /\partial x_\mu$,
$x_\mu~\!\!=~\!\!(x^0=ct,\bmv{x})$; $\gamma_\mu$ are Dirac matrices
\cite{c51,c52,c53}.

The nonequilibrium state of such a system is described by a relativistic
quantum nonequilibrium statistical operator $\varrho (t)$ which
satisfies the relativistic Liouville equation admitting an obviously
covariant form. Such a Liouville equation for a thermal quantum field
system in a covariant form in the interaction representation was 
written down in \cite{c67,c70}  on the basis of the Tomonaga-Schwinger
equation \cite{c71,c72} in the next form:
%
%
\be
\frac{\delta}{\delta\sigma (x)}\varrho_{\rm int}(\sigma)-
\frac 1{\im\hbar}\ls H_{\rm int}(x),\varrho_{\rm int}(\sigma)\rs=0,
\label{e9.2}
\ee
where the nonequilibrium statistical operator $\varrho (\sigma)$ is defined
on an arbitrary space-like surface $\sigma(x)$ \cite{c71,c72,c73}. When the
surface $\sigma (x)$ tends to the plane $t={\rm const}$, equation
(\ref{e9.2}) transforms into the quantum Liouville equation in the
interaction representation,
\bean
\frac{\partial}{\partial t}\varrho_{\rm int}(t)-\frac1{\im\hbar}
\ls H_{\rm int}(x),\varrho_{\rm int}(t)\rs=0,
\eean
where $\varrho_{\rm int}(t)$ is a nonequilibrium statistical operator
in the interaction representation, given on the plane $t={\rm const}$:
\[
H_{\rm int}(t)=\exp\lc-\frac{1}{\im\hbar}H_0t\rc H_{\rm int}
	\exp\lc\frac1{\im\hbar}H_0t\rc,\;\;
	\varrho_{\rm int}(t)=\exp\lc-\frac{1}{\im\hbar}H_0t\rc\varrho(t),
\]
$H_0$ and $H_{\rm int}$ are noninteracting and interacting parts,
correspondingly, of the total Hamiltonian $H$ of the system. The Hamiltonian
of a quark-gluon system corresponding to the Lagrangian (\ref{e9.1}) was
obtained in \cite{c74} using the Coulomb gauge. We will represent it in the 
following form:
%
%
\bea
H&=&H_0+H_{\rm int},\label{e9.3}\\
H_0&=&\frac 12P_l^aP_l^a+\frac 12\partial _nA_l^a\partial _nA_l^a,\nonumber\\
H_{\rm int}&=&gf^{abc}\partial _lA_n^aA_l^bA_n^c+\frac{g^2}2
f^{abc}f^{ade}A_l^bA_n^cA_l^dA_n^e-\frac 12A_0^a\Delta A_0^a+{}\nonumber\\
&&{\mit\Psi}^{+}\left[\left(\im\partial_l+\frac g2\lambda^aA_l^a\right)
\gamma_l+m-\frac g2\lambda^aA_0^a\gamma_0\right]{\mit \Psi},\nonumber
\eea
selecting the ``free field'' Hamiltonian $H_0$ and the one which describes
an interaction between quark and gluon fields. Here $m$ is a colour
independent mass matrix for quark flavour indices, $P_l^a$ means a 
canonical momentum conjugated to gluon field $A^a_l$, while zeroth
components $A_0^a$, like in electrodynamics, are not independent  and
should be inserted in $H_{\rm int}$ after the solution of the equation of 
motion:
\[
\Delta A_0^a=gf^{abc}A_l^b\left(P_l^c+\partial_lA_0^c\right)+
\frac g2{\mit\Psi}^{+}\gamma_0\lambda^a{\mit\Psi}.
\]
On the contrary, proceeding from electrodynamics, it is possible only
in the form of infinite series, so that Hamiltonian (\ref{e9.3}) really
consists of an infinite number of vertices.

To solve the relativistic Liouville equation (\ref{e9.2}), boundary
conditions should be set. Let us search the solutions which depend on time 
through a certain set of observable quantities only, the number of which is 
sufficient for the description of the nonequilibrium state of a system, 
using the method of a nonequilibrium statistical operator. For this 
purpose, we introduce the infinitesimal source
$-\eta\Big(\varrho_{\rm int}(\sigma )-
\varrho^{\rm int}_{\rm q}(\sigma )\Big)$ in the
right-hand side of equation (\ref{e9.2}), which corresponds to the boundary
condition $\varrho_{\rm int}(\sigma)\to\varrho^{\rm int}_{\rm q}(\sigma)$
with $\sigma\to-\infty$ and, according to the formalism of nonequilibrium
thermo field dynamics, write down this equation in the thermo field
representation:
%
%
\be
\ds\frac\delta{\delta\sigma (x)}|\varrho_{\rm int}(\sigma )\ra\ra-
	\frac1{\im\hbar}\bar{H}_{\rm int}(x)|\varrho_{\rm int}(\sigma )\ra\ra=
	-\eta \Big( |\varrho_{\rm int}(\sigma )\ket\ket-
	|\varrho^{\rm int}_{\rm q}(\sigma )\ket\ket\Big),
	\label{e9.4}
\ee
where the source selects retarded solutions with respect to a reduced
description of the nonequilibrium state of a system.
$|\varrho^{\rm int}_{\rm q}(\sigma )\ra\ra$ is a quasi-equilibrium
state-vector; $\bar{H}_{\rm int}(x)$ is the Hamiltonian of interaction
between the quark and gluon superfields,
\bean
\bar H_0(x)=\hat H_0(x)-\tilde H_0(x),\qquad
\bar H_{\rm int}(x)=\hat H_{\rm int}(x)-\tilde H_{\rm int}(x),
\eean
where $\hat{H}_0$ and $\tilde{H}_0$ bilinearly depend on Bose superfields
$\hat{A}^a_\mu$ and $\tilde{A}^a_\mu$ (without and with a tilde line) with
corresponding commutation relations analogous to those for the Bose fields
\cite{c51}; $\hat{H}_{\rm int}$ and $\tilde{H}_{\rm int}$ are Hamiltonians
composed of gluon Bose superfields $\hat{A}^a_\mu$ and $\tilde{A}^a_\mu$, of
quark Fermi superfields $\hat{{\mit \Psi}}^+$ and $\hat{{\mit \Psi}}$
without a tilde line, and $\tilde{{\mit \Psi}}^+$ and $\tilde{{\mit \Psi}}$
with a tilde with corresponding commutation alignments which are similar for
both Bose $A^a_\mu$ and Fermi fields ${\mit \Psi}$ \cite{c51}. We shall
write ``Schr\"odinger'' equation (\ref{e9.4}) in an integral form,
introducing
%
%
\bea
\lefteqn{\ds|\varrho_{\rm int}(\sigma )\ket\ket=
	|\varrho^{\rm int}_{\rm q}(\sigma)\ra\ra-{}}\label{e9.5}\\
\lefteqn{\ds\int\limits_{-\infty}^\sigma\d^4x'\;
	\exp\lp\eta\Omega_{\sigma'\sigma}\rp
	T(\sigma,\sigma')\lp\frac\delta{\delta\sigma'(x')}-\frac1{\im\hbar}
	\bar H_{\rm int}(x')\rp|\varrho^{\rm int}_{\rm q}(\sigma')\ra\ra,}
	\nonumber
\eea
where $\Omega_{\sigma',\sigma}$ means the volume enclosed between the
surface of integration $\sigma'(x')$ and the second one $\sigma (x)$;
$T(\sigma,\sigma')$ is an evolution operator,
\bean
T(\sigma,\sigma')=\exp_+\lc\int\limits_{\sigma'}^\sigma
	\d^4x'\;\frac1{\im\hbar}\bar H_{\rm int}(\sigma')\rc.
\eean
The quasiequilibrium statistical superoperator
$\hat{\varrho}_{\rm q}(\sigma)$ in
$|\varrho_{\rm q}(\sigma)\ra\ra=\hat{\varrho}_{\rm q}(\sigma)|1\ra\ra$
is defined generally \cite{c27,c33,c34} from the entropy extreme with 
keeping the normalization and under the conditions that observable 
quantities $\la p_n\ra^t=\la\la1|\hat{p}_n|\varrho(\sigma)\ra\ra$ are given.
There are the following characteristic values for QGP: mean density of the 
quark barion charge $\la b_\mu (x)\ra^t$, mean densities of the colour
currents $\la j_{{\rm q}\ \mu}^a(x)\ra^t$ for quarks and
$\la j_{{\rm gl}\ \mu}^a(x)\ra^t$ for gluons and also values of the 
densities of the energy-momentum tensor $\la T^{\rm q}_{\mu\nu}(x)\ra^t$ 
for quarks and $\la T^{\rm gl}_{\mu\nu}(x)\ra^t$ for gluons, in which 
densities $b_\mu(x)$, $j_{{\rm q}\ \mu}^a(x)+j_{{\rm gl}\ \mu}^a(x)$,
$T^{\rm q}_{\mu\nu}(x)+T^{\rm gl}_{\mu\nu}(x)$ satisfy the local 
conservation laws,
%
%
\be
\ba{l}
\ds\partial_\mu b_\mu (x)=0,\\[0.5ex]
\ds\partial_\mu\ls j_{{\rm q}\ \mu}^a(x)+j_{{\rm gl}\ \mu}^a(x)\rs
	+gf^{abc}A_\mu^b(x)
	\ls j_{{\rm q}\ \mu}^c(x)+j_{{\rm gl}\ \mu}^c(x)\rs=0,\\[0.5ex]
\ds\partial_\mu\ls T_{\mu\nu}^{\rm q}(x)+T_{\mu\nu}^{\rm gl}(x)\rs+
	g\ls j_{{\rm q}\ \mu}^a(x)+j_{{\rm gl}\ \mu}^a(x)\rs
	F_{\mu\nu}^a(x)=0,\\[0.5ex]
\ds\partial_\mu F_{\mu\nu}^a(x)+gf^{abc}A_\mu^b(x)F_{\mu\nu}^c(x)-
	g\Big(j_{{\rm q}\ \nu}^a(x)+j_{{\rm gl}\ \nu}^a(x)\Big)=0.\\
\ea
\label{e9.6}
\ee
The densities of operators $b_\mu (x)$,
$j_{{\rm q}\ \mu}^a(x)+j^a_{{\rm gl}\ \mu}(x)$,
$T^{\rm q}_{\mu\nu}(x)+T^{\rm gl}_{\mu\nu}(x)$ are ``slowly-changing'' and
their mean values on  the  long time scale satisfy the relativistic
equations of hydrodynamics; it is necessary to co-ordinate with them the
kinetic equations for quark and gluon distribution functions. Local laws
(\ref{e9.6}) impose some restrictions on the kinetic processes, and what
is more, their role is considerably important at large densities and
strong interactions. It shows that the kinetics and hydrodynamics of a 
quark-gluon system are strongly correlated. That is why, in writing down 
the kinetic equations for such systems, it is natural to choose the reduced 
description of the nonequilibrium state in such a way that the proper 
dynamics of conserved quantities is taken into account automatically.

We introduce Wigner operators for quarks and gluons \cite{c50,c64,c65,c75} 
to obtain the kinetic equations:
%
%
\be
f(x;p)=-\int\frac{\d^4y}{\lp2\pi\hbar\rp^4}\exp\lc\frac1{\im\hbar}py\rc
	U\lp x,x_{-}\rp{\mit \Psi}\lp x_{-}\rp{\mit \Psi}^+
	\lp x_{+}\rp U\lp x_{+},x\rp
	\label{e9.7}
\ee
is the Wigner operator of quark density;
%
%
\bea
\ds G_{\mu\nu}(x;p)&=&\int\frac{\d^4y}{\lp2\pi\hbar\rp^4}\exp
	\lc\frac1{\im\hbar}py\rc\times{}\label{e9.8}\\
&&\Big[ U(x,x_-)F_{\mu\lambda}(x_-)U(x_-,x)\Big]\otimes
	\Big[ U(x,x_+)F_\nu^\lambda(x_+)U(x_+,x)\Big]\nonumber
\eea
is the Wigner operator which is connected with the density operator of an 
energy-momentum tensor for gluons:
%
%
\be
T_{\mu\nu}^{\rm gl}(x)=\int\d^4p\;\la\Sp G_{\mu\nu}(x)-\frac 14g_{\mu\nu}
	\Sp G_\lambda^\lambda(x)\ra.
	\label{e9.9}
\ee
Here $U(x',x)$ is an operator which was introduced in paper \cite{c73},
%
%
\be
\ba{llll}
\ds U(x',x)=\exp_+\lc-\frac{g}{\im\hbar c}
	\int\limits_{x}^{x'}\d z^\mu\; A_\mu (z)\rc,\\
\ds x_-=x-y/2,&\hspace*{-8em}
	F_{\mu\nu}&=&\ds F_{\mu\nu}^a(x)\lp\frac{\hbar\lambda_a}{2}\rp,\\
\ds x_+=x+y/2,&\hspace*{-8em}
	A_\mu (x)&=&\ds A_\mu^a (x)\lp\frac{\hbar\lambda_a}{2}\rp.\\
\ea
\label{e9.10}
\ee
$U(x',x)$ is a connection operator, where the integration path
$z(s)=x+s(x'+x)$, $0\leqslant s\leqslant 1$ lies in the plane $t=\const$.
The densities of barion charge $b_\mu (x)$, quark colour current
$j_{{\rm q}\ \mu}^a(x)$ and energy-momentum tensor $T^{\rm q}_{\mu\nu}(x)$
are defined via the Wigner operator of quark density, for example,
%
%
\bea
b_\mu (x)&=&\int\d^4p\;\la\Sp\gamma_\mu f(x;p)\ra,\label{e9.11}\\
j_{{\rm q}\ \mu}^a (x)&=&\int\d^4p\;\la\Sp\frac{\hbar}{2}\lambda_a
	\gamma_\mu f(x;p)\ra,\label{e9.12}\\
T_{\mu\nu}^{\rm q}(x)&=&\int\d^4p\;\la\Sp\gamma_\mu f(x;p)\ra
	P_\nu.\label{e9.13}
\eea
To provide a consistent description of a quark-gluon system, we define the
quasi\-equilibrium statistical superoperator $\hat{\varrho}_{\rm
q}(\sigma)$, $\hat{\varrho}_{\rm q}(\sigma)|1\ra\ra=|\varrho_{\rm
q}(\sigma)\ra\ra$ in a standard way with keeping to the normalization
condition and the following requirements of the averages \cite{c27,c33,c34}:
%
%
\be
\ba{ll}
\ds\la\la 1|\hat{b}_\mu (x)|\varrho (\sigma)\ra\ra,&\qquad
	\ds\la\la 1|\hat{f}(x;p)|\varrho (\sigma)\ra\ra,\\
\ds\la\la 1|\hat{T}_{\mu\nu} (x)|\varrho (\sigma)\ra\ra,&\qquad
	\ds\la\la 1|\hat{G}_{\mu\nu} (x;p)|\varrho (\sigma)\ra\ra.\\
\ea
\label{e9.14}
\ee
Then $\hat{\varrho}_{\rm q}(\sigma)$ takes the form:
%
%
\be
\hat{\varrho}_{\rm q}(\sigma)=\exp\lc -\hat{S}(\sigma)\rc,\label{e9.15}
\ee
where
%
%
\bea
\lefteqn{\ds\hat{S}(\sigma)=\Phi^+(\sigma)+\int\limits_{\sigma}\d\sigma^\mu\;
	\ls P^\nu\hat{T}_{\mu\nu}-\xi\hat{b}_\mu(x)\rs+{}}\label{e9.16}\\
&&\int\limits_{\sigma}d\sigma^\mu\int\d^4p\;
	\ls a(x;p)\hat{f}(x;p)+\omega^\nu(x;p)\hat{G}_{\mu\nu}(x;p)\rs,
	\nonumber
\eea
$\sigma (x)$ is an arbitrary space-like surface passing through point $x$:
$\d\sigma^\mu=\d\sigma n^\mu$ is a surface element vector; $n^\mu$ is a 
normal vector $(\d\sigma^0=\d^3x)$; $P^\nu (x)$, $\xi (x)$, $a(x;p)$,
$\omega^\nu (x;p)$ are the Lagrange factors that are defined from the
self-consistency conditions:
%
%
\be
\ba{lll}
\la\la 1|\hat{b}_\mu(x)|\varrho (\sigma )\ra\ra
	&=&\la\la 1|\hat{b}_\mu(x)|\varrho_q (\sigma )\ra\ra,\\
\la\la 1|\hat{T}_{\mu\nu}(x)|\varrho (\sigma )\ra\ra
	&=&\ds\la\la 1|\hat{T}_{\mu\nu}(x)|\varrho_q (\sigma )\ra\ra,\\
\la\la 1|\hat{f}(x;p)|\varrho (\sigma )\ra\ra
	&=&\la\la 1|\hat{f}(x;p)|\varrho_q (\sigma )\ra\ra,\\
\la\la 1|\hat{G}_{\mu\nu}(x;p)|\varrho (\sigma )\ra\ra
	&=&\la\la 1|\hat{G}_{\mu\nu}(x;p)|\varrho_q (\sigma )\ra\ra,
\ea\label{e9.17}
\ee
$P_\mu=\beta u_\mu$, $\beta=1/T(x)$ denotes inverse invariant temperature,
$u_\mu (x)$ is a local hydrodynamic velocity; $\xi=\beta\mu$, $\mu_a (x)$ is
a local chemical potential of $a$-sort quarks ($a$: $b$, $c$, $d$, $s$, $t$,
$u$). Parameters $a(x;p)$ and $\omega^\nu (x;p)$ are conjugated to
averages $\la\la 1|\hat{f}(x;p)|\varrho(\sigma)\ra\ra$, $\la\la
1|\hat{G}_{\mu\nu}(x;p)|\varrho(\sigma)\ra\ra$.  $\hat{T}_{\mu\nu}(x)$,
$\hat{b}_\mu (x)$, $\hat{f}(x;p)$, $\hat{G}_{\mu\nu}(x;p)$ are
superoperators constructed according to (\ref{e9.7})--(\ref{e9.13}) and the 
thermo field dynamics formalism \cite{c27}, on Bose superfields
$\hat{A}^a_\mu (x)$ and Fermi su\-per\-fields $\hat{{\mit \Psi}}^+$,
$\hat{{\mit \Psi}}$. ${\mit \Phi}^+(\sigma )$ is calculated from the
normalization condition: $\la\la 1|\varrho_q (\sigma )\ra\ra$ equals to 1.
Accordingly, superoperator $\tilde{\varrho}_{\rm q}^+(\sigma)$ is
built on operators $\tilde{T}_{\mu\nu}(x)$, $\tilde{b}_\mu (x)$,
$\tilde{f}(x;p)$, $\tilde{G}_{\mu\nu} (x;p)$ which are expressed via 
Bose superoperators $\tilde{A}^a_\mu(x)$ and Fermi superoperators
$\tilde{{\mit \Psi}}^+$, $\tilde{{\mit \Psi}}$ (with a tilda line). Taking
into consideration the structure of quasiequilibrium statistical
superoperator \refp{e9.15} and \refp{e9.16}, one can write the
nonequilibrium thermo vacuum state vector \refp{e9.5} in the interaction
representation in the following form:
%
%
\bea
\lefteqn{\ds
\ds|\varrho_{\rm int}(\sigma)\ra\ra=|\varrho^{\rm int}_{\rm q}(\sigma)\ra\ra
	-\intl_{-\infty}^{\sigma}\d^4x'\;\e^{\eta\Omega_{\sigma'\sigma}}
	T(\sigma,\sigma')\times{}}\label{e9.18}\\
\lefteqn{\ds
\left.\left|\intl_0^1\d\tau\;\e^{-\tau S^{\rm int}(\sigma')}
	\lc\frac{\delta}{\delta\sigma'(x')}-\frac1{\im\hbar}\bar{H}_{\rm int}
	(x')\rc S^{\rm int}(\sigma')\e^{\tau S^{\rm int}(\sigma')}
	\varrho^{\rm int}_{\rm q}(\sigma')\Ra\!\!\Ra.}\nonumber
\eea
Using the rule \cite{c71,c72} $\frac{\delta}{\delta\sigma(x)}
\int_{\sigma}\d\sigma^\mu (x')\;A_\mu (x')=
\frac{\partial}{\partial x_\mu}A_\mu (x)$, for
$\Big(\delta S^{\rm int}(\sigma')\Big)/\Big(\delta \sigma '(x')\Big)$
we obtain
\bean
\lefteqn{\ds\frac{\delta}{\delta\sigma' (x')}S^{\rm int}(\sigma')=
	\frac{\partial}{\partial x'_\mu}
	\bigg[ P^\nu T_{\mu\nu}^{\rm int}(x')-
	\xi b_\mu^{\rm int}(x')+{}}\\
&&\int\d^4p\; \ls a(x';p)f^{\rm int}(x';p)+\omega^\nu (x';p)
	G_{\mu\nu}^{\rm int}(x';p)\rs\bigg].
\eean
The action of the operator $\ds-\frac1{\im\hbar}H_{\rm int}(x)$ on
$S^{\rm int}(\sigma ')$ reads:
\bean
\lefteqn{\ds\frac1{\im\hbar}H_{\rm int}(x)S^{\rm int}(\sigma')=-
	\int\limits_{\sigma'}\d\sigma^\mu (x'')\;\ls
	P^\nu\dot{T}_{\mu\nu}^{\rm int}(x'')-
	\xi\dot{b}_\mu^{\rm int}(x'')\rs+{}}\\
&&\int\limits_{\sigma'}\d\sigma^\mu (x'')\int\d^4p\;
	\ls a(x'';p)\dot{f}^{\rm int}(x'';p)+
	\omega^\nu (x'';p)\dot{G}_{\mu\nu}^{\rm int}(x'';p)\rs,
\eean
where
\[
\ba{rcl@{\,}rcl}
\dot{b}_\mu^{\rm int}(x'')&=&\ds
	-\frac1{\im\hbar}\ls H_{\rm int}(x'),b_\mu^{\rm int}(x'')\rs,&
\dot{f}^{\rm int}(x'';p)&=&\ds
	-\frac1{\im\hbar}\ls H_{\rm int}(x'),f^{\rm int}(x'';p)\rs ,\\[1.5ex]
\dot{T}_{\mu\nu}^{\rm int}(x'')&=&\ds
	-\frac1{\im\hbar}\ls H_{\rm int}(x'),T_{\mu\nu}^{\rm int}(x'')\rs,&
\dot{G}_{\mu\nu}^{\rm int}(x'';p)&=&\ds
	-\frac1{\im\hbar}\ls H_{\rm int}(x'),G^{\rm int}_{\mu\nu}(x'';p)\rs.
\ea
\]
Using the nonequilibrium thermo vacuum state vector (\ref{e9.18}), we obtain
a system of coupled equations for averages (\ref{e9.14}) in the interaction
representation:
%
%
\bea
p^\mu\frac{\partial}{\partial x_\mu}\la\la 1|\hat{b}_\mu (x)|
	\varrho_{\rm int}(\sigma )\ra\ra=0,\label{e9.19}\\
p^\mu\frac{\partial}{\partial x_\mu}
	\la\la 1|\hat{T}_{\mu\nu}(x)|\varrho_{\rm int}(\sigma )\ra\ra=
	p^\mu\frac{\partial}{\partial x_\mu}\la\la 1
	|\hat{T}_{\mu\nu}(x)|\varrho^{\rm int}_{\rm q}
	(\sigma)\ra\ra+{}\label{e9.20}\\
p^\mu\frac{\partial}{\partial x_\mu}\int\limits_{-\infty}^\sigma\d^4x'\;
	{\rm e}^{\eta\Omega_{\sigma'\sigma}}\la\la 1|\hat{T}_{\mu\nu}(x)
	T(\sigma,\sigma')|\times{}\nonumber\\
\int\limits_{0}^{1}\d\tau\;
	\e^{-\tau S^{\rm int}(\sigma')}
	\lc\frac{\delta}{\delta\sigma'(x')}-\frac1{\im\hbar}H_{\rm int}(x')
	\rc S^{\rm int}(\sigma')
	\e^{\tau S^{\rm int}(\sigma')}\vrhoq^{\rm int}(\sigma')\ra\ra,
	\nonumber
\eea
%
%
%
%
%
\bea
p\cdot D(x)\la\la 1|\hat{f}(x,p)|\varrho_{\rm int}(\sigma )\ra\ra=
	p\cdot D(x)\la\la 1|\hat{f}(x,p)|\varrho^{\rm int}_{\rm q}
	(\sigma)\ra\ra+{}\label{e9.21}\\
p\cdot D(x)\int\limits_{-\infty}^\sigma\d^4x'\;
	{\rm e}^{\eta\Omega_{\sigma'\sigma}}\la\la 1|\hat{f}(x,p)
	T(\sigma,\sigma')|\times{}\nonumber\\
\int\limits_{0}^{1}\d\tau\;
	\e^{-\tau S^{\rm int}(\sigma')}
	\lc\frac{\delta}{\delta\sigma'(x')}-\frac1{\im\hbar}H_{\rm int}(x')
	\rc S^{\rm int}(\sigma')
	\e^{\tau S^{\rm int}(\sigma')}\vrhoq^{\rm int}(\sigma')\ra\ra,
	\nonumber
\eea
%
%
%
%
%
\bea
p\cdot\tilde{D}(x)\la\la 1|\hat{G}_{\mu\nu}(x,p)|
	\varrho_{\rm int}(\sigma )\ra\ra=
	p\cdot\tilde{D}(x)\la\la 1|\hat{G}_{\mu\nu}(x,p)|\varrho^{\rm int}_{\rm q}
	(\sigma)\ra\ra+{}\label{e9.22}\\
p\cdot\tilde{D}(x)\int\limits_{-\infty}^\sigma\d^4x'\;
	{\rm e}^{\eta\Omega_{\sigma'\sigma}}\la\la 1|\hat{G}_{\mu\nu}(x,p)
	T(\sigma,\sigma')|\times{}\nonumber\\
\int\limits_{0}^{1}\d\tau\;
	\e^{-\tau S^{\rm int}(\sigma')}
	\lc\frac{\delta}{\delta\sigma'(x')}-\frac1{\im\hbar}H_{\rm int}(x')
	\rc S^{\rm int}(\sigma')
	\e^{\tau S^{\rm int}(\sigma')}\vrhoq^{\rm int}(\sigma')\ra\ra,
	\nonumber
\eea
while the space-like surface $\sigma(x)$ tends to the plane $t=\const$. Here
$D(x)$ and $\tilde{D}(x)$ are covariant derivatives. They act on the Wigner 
operators of quarks and gluons, correspondingly:
\[
D_\mu(x)f(x,p)\equiv\lp\dd{x_\mu}+\frac1{g}{\im\hbar c}A_\mu(x)\rp f(x,p),
\]
where $A_\mu(x)=A_\mu^a(x)T_a$ with $\lp T_a\rp_{bc}=-\im\hbar f_{abc}$ is
an $8\times8$ matrix. The tildian covariant derivative acts on the Wigner
operator of gluons
\[
\tilde{D}_\alpha(x)G_{\mu\nu}(x,p)\equiv\dd{x_\alpha}G_{\mu\nu}(x,p)
	\frac{g}{\im\hbar c}\ls A_\alpha,G_{\mu\nu}(x,p)\rs
\]
with a commutator between two $8\times8$ matrices $\lp
A_\alpha\rp^{mn}=A_\alpha^a\lp T_a\rp^{mn}$ and
$\lp\hat{G}_{\mu\nu}\rp^{ab}$. An action of these covariant derivatives on
the Wigner operator of quarks $p\cdot D(x)f(x,p)$ and on the Wigner 
operator of gluons $p\cdot\tilde{D}(x)\hat{G}_{\mu\nu}(x,p)$ was calculated 
in papers \cite{c64,c65}. The set of equations \refp{e9.19}--\refp{e9.22}
without taking into account the barion charge transport
$\la\la1|\hat{b}_\mu(x)|\vrho(t)\ra\ra$ and the total momentum tensor
$\la\la1|\hat{T}_\mu(x)|\vrho(t)\ra\ra$ turns to a connected system of 
kinetic equations for the average values of Wigner operators of quarks 
$\la\la1|\hat{f}(x,p)|\vrho(t)\ra\ra$, gluons 
$\la\la1|\hat{G}_\mu(x,p)|\vrho(t)\ra\ra$. It generalizes the results of
papers \cite{c64,c65,c66}.

We have obtained a system of coupled relativistic transport equations of
a consistent description of the kinetics and hydrodynamics for QGP in 
thermo field representation. This system of equations is strongly nonlinear
and could be used to describe both strongly and weakly nonequilibrium
states of a system. However, it should be noted that the following
transformations for the use of the system of relativistic transport
equations (\ref{e9.19})--(\ref{e9.22}) have to be made. Since the
quasiequilibrium thermo vacuum state vector $|\varrho_{\rm q}(\sigma
)\ra\ra$, which is used in ave\-raging equations
(\ref{e9.19})--(\ref{e9.22}), is not a vacuum state vector for Bose
superfields $\tilde{A}^a_{\mu}$, $\hat{A}^a_{\mu}$ of gluons and Fermi
superfields $\hat{{\mit \Psi}}$, $\hat{{\mit \Psi}}^+$, $\tilde{{\mit
\Psi}}$, $\tilde{{\mit \Psi}}^+$ of quarks, it is necessary to construct
Bose superfields of gluons and Fermi superfields of quarks for which it is
a vacuum state vector, as in paper \cite{c27}. Such superfields will
depend on the parameters of state $P^\nu (x)$, $\beta\mu$, $a(x;p)$,
$\omega^\nu (x;p)$, besides, all the superoperators $\hat{T}_{\mu\nu}(x)$,
$\hat{b}_\mu (x)$, $\hat{f}(x;p)$,
$\hat{G}_{\mu\nu}(x;p)$, $\bar{H}_{\rm int}(x)$, $T(\sigma ,\sigma ')$ in a
system of relativistic transport equations (\ref{e9.19})--(\ref{e9.22})
should be expressed in a such way where $|\varrho_{\rm q}(\sigma)\ra\ra$
is the vacuum state vector. Besides, the use of space-like surface
$\sigma (x)$ is justified at the calculation of generalized equations
in the invariant form. Since the final results do not depend on the
choice of surface $\sigma$, one has to rewrite equations
(\ref{e9.19})--(\ref{e9.22}) on the surface directed to the plane
$t=\const$. So, it is useful to choose $\eta=\varepsilon V^{-1}$, where $V$
denotes the volume occupied by system \cite{c70}. Then
$\Omega_{\sigma'\sigma}=(t'-t)V$ and
$\eta\Omega_{\sigma '\sigma}=\varepsilon(t'-t)$. Thus, in transport
equations (\ref{e9.19})--(\ref{e9.22}),
$\int\limits^{\sigma}_{-\infty}\d^4x'\;\exp\Big(\eta\Omega_{\sigma '\sigma}
\Big)\to\int\limits_{-\infty}^{t}\d t'\;\exp\Big(\varepsilon(t'-t)\Big)$ and
superoperators $\hat{\varrho}_{\rm int}(t)$,
$\hat{\varrho}^{\rm int}_{\rm q}(t)$,
$T(t,t')$ will be given on the plane $t=\const$. Such questions will be
considered in the next paper, in particular, in the investigation of the
relativistic transport equations for weakly nonequilibrium QGP.

\section{Conclusions}

In this paper the conception of nonequilibrium thermo field dynamics on the
basis of a nonequilibrium statistical operator has been applied to the 
construction of transport equations of dense quantum systems. Hydrodynamic 
equations in thermo field representation have been obtained for both 
strongly and weakly nonequilibrium processes. Transport cores, connected 
with transport coefficients of viscosity and thermal conductivity, have been 
defined, too.  They are calculated with the help of a quasiequilibrium 
thermo vacuum state vector. The last one is a vacuum state vector for 
annihilation and creation superoperators which depend on thermodynamic 
parameters $F_n(t)$ of a system. The consistency of both annihilation and 
creation superoperators and the corresponding vacuum state is accomplished 
then by the mentioned approach. In particular, it is achieved by the 
generalization of the Bogolubov transformations in thermo field dynamics 
\cite{c32} for a nonequilibrium case. General transport equations within 
{\sl nonequilibrium thermo field dynamics -- nonequilibrium statistical 
operator} made it possible to obtain an equation of a generalized 
description of the kinetics and hydrodynamics for a dense quantum system 
with strongly coupled states. In such a case, $f_{A\alpha}(x;t)$ -- the 
Wigner function of the $A$-particle coupled state, and the average value of 
the total energy density operator $\la\la1|\hat{H}(\bmv{r})|\vrho(t)\ra\ra$ 
have been chosen as parameters of a shortened description. Particle 
annihilation and creation superoperators for the quasiequilibrium thermo 
vacuum state vector of noninteracting particles were defined in \refp{e8.22}. 
It gives us the possibility for the construction of a diagram technique in 
the calculation of the corresponding transport cores. These investigations 
and calculations are important for actual nuclear systems \cite{c8,c9}.
\pagebreak

We have  considered a method for obtaining generalized transport equations
for QGP -- one of the nuclear matter states. These equations were obtained 
in the most general form. To be used, their structure needs a lot of
transformations, especially for transport cores. The consistent description
of the kinetics and hydrodynamics is based on a set of fundamental 
parameters of a shortened description: the average values of Wigner 
operators for quarks and gluons $\la\la1|\hat{f}(x)|\vrho(t)\ra\ra$,
$\la\la1|\hat{G}_{\mu\nu}(x)|\vrho(t)\ra\ra$ and the average values of the 
density operator of barion charge $\la\la1|\hat{b}_\mu(x)|\vrho(t)\ra\ra$ 
and the total energy-momentum tensor 
$\la\la1|\hat{T}_{\mu\nu}(x)|\vrho(t)\ra\ra$ of quarks and gluons which 
constitute the basis for the hydrodynamic description. The set of equations 
obtained permits the investigations of weakly nonequilibrium processes and 
kinetic equations like the Boltzmann-Vlasov or Lenard-Balescu ones for 
diluted QGP.

The problem of the investigation of transport coefficients: viscosity and 
thermal conductivity as well as excitations of QGP still remains. It might 
be considered in view of the obtained transport equations. This will be the
subject of our future work.

\label{last@page}
\end{document}